%% file: main.tex
\documentclass[aps,prd,reprint,preprintnumbers,groupedaddress,nofootinbib]{revtex4-2}
\input{structure.tex}

\begin{document}

\title{Changing patterns in electroweak precision with new color-charged states:\\ Oblique corrections and the \emph{W} boson mass}

\author{Linda M. Carpenter}
\email{lmc@physics.osu.edu}
\author{Taylor Murphy}
\email{murphy.1573@osu.edu}
\author{Matthew J. Smylie}
\email{smylie.8@osu.edu}
\affiliation{Center for Cosmology and Astroparticle Physics (CCAPP)\\ and Department of Physics, The Ohio State University\\ 191 W. Woodruff Ave., Columbus, OH 43212, U.S.A.}

\date{\today}

\input{TeX/abstract}

\maketitle

%-----

%\title{\textbf{Changing patterns in electroweak precision\linebreak with new color-charged states:\linebreak Oblique corrections and the \emph{W} boson mass}}
%\author[a,b]{Linda M. Carpenter\thanks{\href{mailto:lmc@physics.osu.edu}{lmc@physics.osu.edu}}}
%\author[a,b]{Taylor Murphy\thanks{\href{mailto:murphy.1573@osu.edu}{murphy.1573@osu.edu}}}
%\author[a,b]{Matthew J. Smylie\thanks{\href{mailto:smylie.8@osu.edu}{smylie.8@osu.edu}}}
%\affil[a]{\emph{Department of Physics, The Ohio State University\linebreak
%191 W. Woodruff Ave., Columbus, OH 43210, U.S.A.}}
%\affil[b]{\emph{Center for Cosmology and Astroparticle Physics (CCAPP),\linebreak The Ohio State %University\linebreak
%191 West Woodruff Avenue, Columbus, OH 43210, U.S.A.}}

%\date{\vspace{-7ex}}

%\maketitle

%\input{TeX/abstract}

%\vfill

%\pagebreak

\input{TeX/1_Intro}
\input{TeX/2_STU}
\input{TeX/3_Model}
\input{TeX/4_Numbers}
\input{TeX/5_Conclusions}

\acknowledgments

This work was supported by the Department of Physics of The Ohio State University.

\bibliographystyle{apsrev4-2}
\bibliography{Bibliography/bibliography.bib}

\end{document}

%% file: structure.tex
\usepackage[svgnames]{xcolor} % Must load first for [svgnames]

\usepackage[printonlyused,withpage]{acronym} % Include a list of acronyms
\usepackage{amsfonts}
\usepackage{anyfontsize}
\usepackage{bbm}
\usepackage{booktabs} % Required for better horizontal rules in tables
\usepackage{braket}
\usepackage{cancel}
\usepackage{changepage}
\usepackage{colortbl}
\usepackage{enumitem}
\usepackage[mathscr]{euscript}
\usepackage{float}
\usepackage[T1]{fontenc} % Support for more character glyphs
\usepackage{gensymb}
\usepackage[hidelinks]{hyperref} % Handles links and changing link options
\usepackage{listings} % Required for including snippets of code
\usepackage{makecell}
\usepackage{mathtools}
\usepackage[framemethod=tikz]{mdframed} % Fancy framing and shading
\usepackage{microtype} % Slightly tweak font spacing for aesthetics
\usepackage{multirow}
\usepackage{physics}
\usepackage{relsize}
\usepackage{rotating} % Allows tables and figures to be rotated
\usepackage{sansmath}
\usepackage{simplewick}
\usepackage{slashed}
\usepackage{stmaryrd}
\usepackage{subfig}
\usepackage{tcolorbox}
\usepackage{theorem}
\usepackage{tikz}
\usepackage{tikz-feynman,contour}
\usepackage[normalem]{ulem}
\usepackage{xspace} % Intelligent spacing for \institution and \department
\usepackage{yfonts}
\usepackage[vcentermath]{youngtab}

\usepackage{newtxtext,newtxmath}

\tcbuselibrary{theorems}
\usetikzlibrary{decorations.markings}
\usetikzlibrary{shapes.geometric}

\def\ii{{\text{i}}}

\makeatletter
\newcommand*{\transpose}{%
  {\mathpalette\@transpose{}}%
}
\newcommand*{\@transpose}[2]{%
  \raisebox{\depth}{$\m@th#1\intercal$}%
}
\makeatother

\newtcbox{\sln}{colback=Gainsboro,
colframe=Gainsboro}

\newcommand{\bt}[1]{{\sansmath{\boldsymbol{#1}}}}

\tikzset{snake it/.style={decorate, decoration={snake,amplitude=10mm}}}

\tikzset{/pgf/decoration/.cd,
    number of sines/.initial=10,
    angle step/.initial=20,
}

\newdimen\tmpdimen

\pgfdeclaredecoration{full sines}{initial}
{
    \state{initial}[
        width=+0pt,
        next state=move,
        persistent precomputation={
            \pgfmathparse{\pgfkeysvalueof{/pgf/decoration/angle step}}%
            \let\anglestep=\pgfmathresult%
            \let\currentangle=\pgfmathresult%
            \pgfmathsetlengthmacro{\pointsperanglestep}%
                {(\pgfdecoratedremainingdistance/\pgfkeysvalueof{/pgf/decoration/number of sines})/360*\anglestep}%
        }] {}
    \state{move}[width=+\pointsperanglestep, next state=draw]{
        \pgfpathmoveto{\pgfpointorigin}
    }
    \state{draw}[width=+\pointsperanglestep, switch if less than=1.25*\pointsperanglestep to final, % <- bit of a hack
        persistent postcomputation={
        \pgfmathparse{mod(\currentangle+\anglestep, 360)}%
        \let\currentangle=\pgfmathresult%
    }]{%
        \pgfmathsin{+\currentangle}%
        \tmpdimen=\pgfdecorationsegmentamplitude%
        \tmpdimen=\pgfmathresult\tmpdimen%
        \divide\tmpdimen by2\relax%
        \pgfpathlineto{\pgfqpoint{0pt}{\tmpdimen}}%
    }
    \state{final}{
        \ifdim\pgfdecoratedremainingdistance>0pt\relax
            \pgfpathlineto{\pgfpointdecoratedpathlast}
        \fi
   }
}

\pgfdeclaredecoration{completely sines}{initial}
{
    \state{initial}[
        width=+0pt,
        next state=upsine,
        persistent precomputation={\pgfmathsetmacro\matchinglength{
            \pgfdecoratedinputsegmentlength / int(\pgfdecoratedinputsegmentlength/\pgfdecorationsegmentlength)}
            \setlength{\pgfdecorationsegmentlength}{\matchinglength pt}
        }] {}
    \state{upsine}[width=\pgfdecorationsegmentlength,next state=downsine]{
        \pgfpathsine{\pgfpoint{0.25\pgfdecorationsegmentlength}{0.5\pgfdecorationsegmentamplitude}}
        \pgfpathcosine{\pgfpoint{0.25\pgfdecorationsegmentlength}{-0.5\pgfdecorationsegmentamplitude}}
    }
    \state{downsine}[width=\pgfdecorationsegmentlength,next state=upsine]{
        \pgfpathsine{\pgfpoint{0.25\pgfdecorationsegmentlength}{-0.5\pgfdecorationsegmentamplitude}}
        \pgfpathcosine{\pgfpoint{0.25\pgfdecorationsegmentlength}{0.5\pgfdecorationsegmentamplitude}}
}
    \state{final}{}
}

\tikzfeynmanset{ with arrow/.style = {
   decoration={
     markings,
     mark=at position 0.5
          with {\arrow[xshift=1.9mm]{Latex[width=2.75mm,length=3mm]}}
     },
   postaction=decorate}
}

\tikzfeynmanset{ with glarrow/.style = {
   decoration={
     markings,
     mark=at position 0.5
          with {\arrow[white,xshift=3mm]{Latex[width=4.125mm,length=4.5mm]}}
     },
   postaction=decorate}
}

\tikzfeynmanset{ with outarrow/.style = {
   decoration={
     markings,
     mark=at position 0.5
          with {\arrow[xshift=3.35mm]{Latex[width=4.58333mm,length=5mm]}}
     },
   postaction=decorate}
}

\tikzfeynmanset{ bigphoton/.style={
    /tikz/draw=none,
    /tikz/decoration={name=none},
    /tikz/postaction={
      /tikz/draw,
      /tikz/decoration={
        completely sines,
        segment length=3mm,
        amplitude=2.5mm,
      },
      /tikz/decorate=true,
    },
  },
}

\tikzfeynmanset{ roundbigphoton/.style={
    /tikz/draw=none,
    /tikz/decoration={name=none},
    /tikz/postaction={
      /tikz/draw,
      /tikz/decoration={
        full sines,
        segment length=3mm,
        amplitude=1.25mm,
      },
      /tikz/decorate=true,
    },
  },
}

\tikzset{ mega thick/.style= {line width = 3.4pt}
}

\makeatletter
\renewcommand{\fnum@figure}{\textsc{\figurename~\thefigure}} % Make the "Figure 1.1" text in small caps
\makeatother

\lstnewenvironment{pseudoc}{\lstset{frame=lines,language=C,mathescape=true}}{}
\lstnewenvironment{logs}{\lstset{frame=lines,basicstyle=\footnotesize\ttfamily,numbers=none}}{}
\lstnewenvironment{cc}{\lstset{frame=lines,language=C}}{}
\lstnewenvironment{c64}{\lstset{backgroundcolor=\color{c64},basewidth=0.65em,basicstyle=\commodoreface\color{c64light},numbers=none,framerule=10pt,rulecolor=\color{c64light},frame=tb,framexbottommargin=30pt}}{}
\lstnewenvironment{html}{\lstset{frame=lines,language=html,numbers=none}}{}
\lstnewenvironment{pseudo}{\lstset{frame=lines,mathescape=true,morekeywords={learn_string_domain, save_model}}}{}
\lstnewenvironment{pseudoctiny}{\lstset{language=C,mathescape=true,basicstyle=\tiny\sffamily}}{}
\lstnewenvironment{cctiny}{\lstset{language=C,basicstyle=\tiny\sffamily}}{}
\lstnewenvironment{pseudotiny}{\lstset{mathescape=true,basicstyle=\tiny\sffamily}}{}

%% file: TeX/abstract.tex
\begin{abstract}
    
The recent measurement by the CDF Collaboration of the $W$ boson mass is in significant tension with the Standard Model expectation, showing a discrepancy of seven standard deviations. A larger value of $m_W$ affects the global electroweak fit, particularly the best-fit values of the Peskin-Takeuchi parameters $S$, $T$ (and perhaps $U$) that measure oblique corrections from new physics. To meet this challenge, we propose some simple models capable of generating non-negative $S$ and $T$, the latter of which faces the greatest upward pressure from the CDF measurement in scenarios with $U=0$. Our models feature weak multiplets of scalars charged under $\mathrm{SU}(3)_{\text{c}} \times \mathrm{U}(1)_Y$, which cannot attain nonzero vacuum expectation values but nevertheless produce \emph{e.g.} $T \neq 0$ given some other mechanism to split the electrically charged and neutral scalars. We compute the oblique corrections in these models and identify ample parameter space supporting the CDF value of $m_W$.
    
\end{abstract}

%% file: TeX/1_Intro.tex
\section{Introduction}
\label{s1}

The CDF Collaboration recently reported a measurement of the $W$ boson mass based on the full $\mathcal{L}=8.8\,\text{fb}^{-1}$ dataset of proton-antiproton ($p\bar{p}$) collisions collected by the CDF II detector at the Fermilab Tevatron collider \cite{CDFWboson}. CDF reports a mass of
\begin{align}\label{CDFmeas}
    m_W^{\text{CDF II}} = 80.4335 \pm 0.0094\,\text{GeV},
\end{align}
with uncertainties dominated by modeling of the $W$ transverse mass and the transverse momenta of the leptonic products $\ell,\, \nu_{\ell}$ of $W$ decay within the detector. The central value of this measurement is roughly $75\,\text{MeV}$, or $0.1\%$, higher than the extant prediction for $m_W$ within the Standard Model (SM) given by the global fit of electroweak precision observables \cite{Haller:2018nnx,pdg2020},
\begin{align}\label{SMpred}
    m_W^{\text{SM}} = 80.359 \pm 0.006\,\text{GeV}.
\end{align}
The $\mathcal{O}(10^{-4})$ precision of the SM prediction far exceeds those of previous measurements of $m_W$ at the Tevatron \cite{CDF:2013dpa}, LEP, ATLAS \cite{ATLAS:2017rzl}, and LHCb \cite{LHCb:2021bjt}; the new CDF measurement is the first to approach precision parity with theory. Strikingly, this higher precision is not accompanied by convergence with the SM prediction; instead, \eqref{CDFmeas} exceeds \eqref{SMpred} by seven standard deviations ($7\sigma$) assuming independent uncertainties. We therefore face the sudden and urgent question of what new physics can accommodate a higher value of $m_W$.

An impressive amount of activity has already been generated by the CDF measurement, seeking variously to perform the global electroweak fit with new input for $m_W$ and to propose a panoply of models --- supersymmetric theories, neutrino seesaw models, doublet and triplet extensions of the Higgs sector, and others --- to produce a heavier $W$ boson \cite{Kanemura:2022ahw,
Nagao:2022oin,
Perez:2022uil,
Ghoshal:2022vzo,
Kawamura:2022uft,
Zheng:2022irz,
Han:2022juu,
Ahn:2022xeq,
Balkin:2022glu,Biekotter:2022abc,
Endo:2022kiw,
Crivellin:2022fdf,
Cheung:2022zsb,
Du:2022brr,
Heo:2022dey,
Babu:2022pdn,
Heckman:2022the,
Gu:2022htv,
Athron:2022isz,
DiLuzio:2022xns,
Asadi:2022xiy,
Bahl:2022xzi,
Paul:2022dds,
Bagnaschi:2022whn,
Song:2022xts,
Cheng:2022jyi,
Lee:2022nqz,
Liu:2022jdq,
Fan:2022yly,
Sakurai:2022hwh,
Fan:2022dck,
Lu:2022bgw,
Athron:2022qpo,
Yuan:2022cpw,
Strumia:2022qkt,
Yang:2022gvz,
deBlas:2022hdk}. Many of these works have viewed this problem through the lens of the \emph{Peskin-Takeuchi} parameters $S$, $T$, and $U$, which can be used to measure the size of certain simple kinds of novel electroweak phenomena (\emph{oblique corrections}) with the potential to affect the global electroweak fit \cite{Peskin:1990zt,Peskin:1991sw}. Analyses of the global fit performed since the announcement of the CDF measurement have found that a larger value of $m_W$ (in particular, a larger value relative to the well measured mass of the $Z$ boson) causes the best-fit values of $S$, $T$, and perhaps $U$ to drift above their Standard Model values, which vanish by construction. From this point of view, the challenge to produce a heavier $W$ can therefore be recast as a quest to generate sizable oblique corrections --- preferably large positive values of $T$, it turns out --- in models with new electroweak field content.
  
In this work, we propose to study the oblique corrections generated in a class of models featuring exotic scalar fields carrying both $\mathrm{SU}(2)_{\text{L}}$ weak and $\mathrm{SU}(3)_{\text{c}}$ color charge. New scalars transforming in various representations of $\mathrm{SU}(3)_{\text{c}}$ have garnered considerable theoretical attention in many contexts. Likely the most familiar are scalars in the fundamental representation of $\mathrm{SU}(3)_{\text{c}}$ appearing in models of leptoquarks \cite{Crivellin:2020ukd} --- in which they can also carry weak quantum numbers --- or as squarks in supersymmetric scenarios. Also thoroughly studied are electroweak-singlet color-octet scalars, whose collider phenomenology has been explored, for example, in \cite{Gerbush:2007fe,Choi:2008ub,Plehn:2008ae,Chen:2014haa,Carpenter:2015gua,Carpenter:2020hyz,Carpenter:2020evo,Carpenter:2021vga}. Color-sextet fields have enjoyed less attention \cite{Chen:2008hh,Han:2010rf}, but have seen renewed interest of late \cite{Carpenter:2021rkl}. Models with scalars in higher-dimensional representations of $\mathrm{SU}(3)_{\text{c}}$ have several potential benefits relevant to the immediate goal of generating sizable one-loop electroweak corrections. The first is simply that larger color factors tend to enhance loop corrections relative to those from new colorless states. The other is that states in higher-dimensional representations of $\mathrm{SU}(3)_{\text{c}}$ that also carry $\mathrm{SU}(2)_{\text{L}}$ (and optionally $\mathrm{U}(1)_Y$) charge are less constrained by collider searches than, for example, states with leptoquark quantum numbers.

We specifically consider two model frameworks: in the first, the Standard Model is extended by two $\mathrm{SU}(2)_{\text{L}}$ doublets with identical $\mathrm{SU}(3)_{\text{c}} \times \mathrm{U}(1)_Y$ quantum numbers, which mix with each other and generate splitting between electrically charged and neutral color-charged scalars --- hence, in principle, at least non-vanishing $T$ --- even if sequestered from the SM Higgs doublet. In the second scenario, we introduce a multiplet with vanishing weak hypercharge transforming in the adjoint representations of both $\mathrm{SU}(3)_{\text{c}}$ and $\mathrm{SU}(2)_{\text{L}}$. This \emph{biadjoint model} ultimately requires an additional color-octet field to generate splitting between the charged and neutral scalars. We compute the potentially sizable oblique corrections, as measured by $S$ and $T$, in both frameworks, and we reconcile the parameter space with direct LHC search limits for similar new scalars.
  
This paper is organized as follows. In \hyperref[s2]{Section II}, we briefly review the global electroweak fit in light of the CDF measurement of $m_W$, and we discuss the parametrization of new electroweak physics offered by the Peskin-Takeuchi parameters $S,\,T,\,U$. We review the common choice to set $U=0$ and discuss the ranges of $S$ and $T$ we intend to target in our numerical investigation. In \hyperref[s3]{Section III}, we introduce our simple models with novel field content carrying $\mathrm{SU}(3)_{\text{c}} \times \mathrm{SU}(2)_{\text{L}} \times \mathrm{U}(1)_Y$ quantum numbers. We make some general remarks about these models, including about their mass spectra and the allowed multiplicity of color-charged scalars, and compute the oblique corrections. With an eye toward collider constraints, we write an effective field theory coupling these weak multiplets to Standard Model gauge bosons. \hyperref[s4]{Section IV} explores the parameter space of our models, both with respect to $S$ and $T$ and in view of possible constraints from the Large Hadron Collider. \hyperref[s5]{Section V} concludes.  

%% file: TeX/2_STU.tex
\section{Oblique corrections and the \emph{W} mass}
\label{s2}

\renewcommand\arraystretch{2}
\begin{table}\label{EWfit}
    \centering
    \begin{tabular}{|c|| c |}
    \toprule
\hline 
Observable & Measured Value\\ 
\hline
\hline
$m_Z\,\text{[GeV]}$ & $91.1875 \pm 0.0021$\\
\hline
$\sin^2 \theta^{\ell}_{\text{eff}}$(Tevt.) & \ $0.23148 \pm 0.00033$\ \ \\
\hline
$\alpha^{-1}(m_e^2)$ & 137.03599084\\
\midrule
$m_W^{\text{SM}}\,\text{[GeV]}$ & $80.359 \pm 0.006$\\
\hline
\ $m_W^{\text{CDF II}}\,\text{[GeV]}$\ \ & $80.4335 \pm 0.0094$\\
\hline 
\bottomrule
\end{tabular} 
    \caption{Input values used for numerical results in \hyperref[s4]{Section IV}, corresponding to a small subset of values used in the global electroweak fit. Comprehensive list of inputs used by the Gfitter group is available at \cite{Haller:2018nnx}.}
\end{table}
\renewcommand\arraystretch{1}

We begin by reviewing the connections between the mass of the $W$ boson and physics beyond the Standard Model. We first discuss the global electroweak fit of the Standard Model, which has long been used to constrain new physics and has been already been updated by several groups following the CDF measurement of $m_W$. We then explore the paradigm in which new electroweak physics is dominated by \emph{oblique corrections} to SM gauge boson self-energies (vacuum polarizations) and review the Peskin-Takeuchi parametrization of such corrections. We finally summarize the collection of (extremely) recent constraints on oblique corrections in concordance with old and new measurements of $m_W$, identifying the parameter space to be targeted by our models.

The global electroweak fit is a statistical analysis in which a $\chi^2$ test statistic is minimized for a set of $\mathcal{O}(10)$ electroweak precision observables. The current standard analysis strategy was formulated by the Gfitter group, developers of the eponymous \textsc{Gfitter} package \cite{Flacher:2008zq,Baak:2011ze,Baak:2014ora,Haller:2018nnx}, and defines
\begin{align}
    \chi^2(y_{\text{mod}}) \equiv -2 \ln \mathcal{L}(y_{\text{mod}})
\end{align}
with $\mathcal{L}(y_{\text{mod}})$ the product of theoretical and experimental likelihoods for some model parameters $y_{\text{mod}}$ that are ideally Gaussian and independent but in reality may or may not be correlated \cite{Flacher:2008zq}. Various implementations of the global fit have been used since before the discovery of the top quark to validate the electroweak Standard Model and impose constraints on new electroweak physics. The power of the global electroweak fit in both arenas has grown with the remarkable improvements in precision of experimental results and theoretical predictions in the last few decades.\footnote{Validation of the Standard Model with the global fit is possible now because the model has been overconstrained since the measurement of the SM Higgs mass \cite{Haller:2018nnx}.} The global fit takes inputs from observables directly related to weak bosons --- including the mass of the $W$ boson --- and the Higgs, along with an array of observables parametrizing the interactions of weak bosons with quarks and leptons. The precision measurements themselves come from the Stanford Linear Collider (SLC), the Large Electron-Positron Collider (LEP) at CERN, the Tevatron, and the Large Hadron Collider. Predictions are made for many of the precision observables within the Standard Model, at up to two-loop order, using a subset of the measured free parameters of the model. Some of the many inputs currently used in the global fit are displayed in \hyperref[EWfit]{Table I}, which we use in our quantitative investigation in \hyperref[s4]{Section IV}. The global fit has for many years found good agreement between the Standard Model and the electroweak precision data taken as a whole --- though, clearly, the situation may be different now --- and has been used to constrain or outright forbid a host of scenarios beyond the Standard Model in which electroweak physics is measurably altered.

One well known and well motivated class of new electroweak physics induces radiative corrections to the electroweak gauge boson self-energies, potentially affecting the masses and mixing of these bosons, but does not alter the SM interactions between these bosons and the quarks or leptons. This restricted class of models, limited to so-called oblique corrections, can be directly probed by four-fermion scattering at lepton and hadron colliders and can also be constrained by the global electroweak fit. Under certain simplifying assumptions, oblique corrections from new physics can be parametrized by the Peskin-Takeuchi parameters $S$, $T$, and $U$ \cite{Peskin:1990zt,Peskin:1991sw}, which can be written in terms of electroweak gauge boson self-energies $\Pi_{XY}(p^2)$, $X,Y \in \{W,Z,\gamma\}$, as
\begin{widetext}
\begin{align}\label{PTparam}
\nonumber     S &= \frac{4}{\alpha}\,(s_{\text{w}}c_{\text{w}})^2\,\frac{1}{m_Z^2}\left[\Pi_{ZZ}^{\text{new}}(m_Z^2) - \Pi_{ZZ}^{\text{new}}(0) - \frac{c_{\text{w}}^2-s_{\text{w}}^2}{s_{\text{w}}c_{\text{w}}}\, \Pi_{Z\gamma}^{\text{new}}(m_Z^2) - \Pi_{\gamma\gamma}^{\text{new}}(m_Z^2)\right],\\
\nonumber     T &= \frac{1}{\alpha}\left[\frac{\Pi_{WW}^{\text{new}}(0)}{m_W^2} - \frac{\Pi_{ZZ}^{\text{new}}(0)}{m_Z^2}\right] = \frac{1}{\alpha}\,(\rho-1),\\
\text{and}\ \ \ U &= \frac{4}{\alpha}\,s_{\text{w}}^2\left[\frac{\Pi_{WW}^{\text{new}}(m_W^2)- \Pi_{WW}^{\text{new}}(0)}{m_W^2} - \frac{c_{\text{w}}}{s_{\text{w}}}\frac{\Pi_{Z\gamma}^{\text{new}}(m_Z^2)}{m_Z^2} - \frac{\Pi_{\gamma\gamma}^{\text{new}}(m_Z^2)}{m_Z^2}\right]- S,
\end{align}
\end{widetext}
where $s_{\text{w}} = \sin \theta_{\text{w}},\ c_{\text{w}} = \cos \theta_{\text{w}}$ are functions of the weak mixing angle $\theta_{\text{w}}$, and where
\begin{align}\label{rhodef}
    \rho = \frac{m_W^2}{m_Z^2 c_{\text{w}}^2} = 1 + \delta \rho
\end{align}
measures the violation of the SM $\mathrm{SU}(2)_V$ custodial symmetry. These parameters are a reliable bellwether of new physics provided that (a) by construction, non-oblique corrections are suppressed or made to vanish, (b) the electroweak gauge group $\mathrm{SU}(2)_{\text{L}} \times \mathrm{U}(1)_Y$ is not extended, and (c) the characteristic energy scale of new physics is not too close to the $Z$ mass. Within this regime of validity, the constraints imposed by the global electroweak fit on physics beyond the Standard Model can be translated into constraints on $S$, $T$, and $U$. If the first two criteria are satisfied, but the third is violated, then an additional trio of parameters often denoted by $V,W,X$ is required to fully parametrize new-physics oblique corrections \cite{Burgess:1993mg}. We neglect these parameters for simplicity in this work. To make further contact with the literature, we note that the traditional parameters $S,\,T,\,U$ correspond to certain operators in the Warsaw basis of the Standard Model Effective Field Theory (SMEFT) \cite{Grzadkowski_2010,Brivio_2017,Murphy_2020}. In particular, we have \cite{Han:2004az,Han:2008es}
\begin{align}
    S = \frac{4}{\alpha}\, s_{\text{w}}c_{\text{w}} v^2\, c_{WB},\ \ \ T = -\frac{v^2}{2\alpha}\, c_h,\ \ \ \text{and}\ \ \ U \sim \frac{4}{\alpha}\, s_{\text{w}} v^4\, c_{Wh},
\end{align}
where the coefficients $c_{WB}$, $c_h$, $c_{Wh}$ are the Wilson coefficients of the operators
\begin{multline}\label{SMEFT}
    \mathcal{L}_{\text{SMEFT}} \supset \frac{c_{WB}}{\Lambda^2}\, (H^{\dagger} \sigma^A H)\,W^A_{\mu\nu}B^{\mu\nu}\\+ \frac{c_h}{\Lambda^2}\, |H^{\dagger} D_{\mu}H|^2\\ + \frac{c_{Wh}}{\Lambda^4}\,(H^{\dagger} W^A_{\mu\nu} H)(H^{\dagger}W^{A\,\mu\nu}H)
\end{multline}
with $\mathrm{SU}(2)_{\text{L}}$ fundamental indices implicit and contracted within parentheses, and the corresponding adjoint indices $A \in \{1,2,3\}$ kept explicit. This representation of the Peskin-Takeuchi parameters is instructive: it shows that $U$ is parametrically suppressed compared to $S$ and $T$, being a dimension-eight operator next to a pair of dimension-six operators \cite{GRINSTEIN1991326}. This fact, and the difficulty of generating large values of $U$ in perturbative models \cite{pdg2020}, frequently motivates the choice to fix $U=0$ in the global electroweak fit. We adopt this scheme in this work, but it is worth noting that $U$ is uniquely sensitive to the mass and decay width of the $W$ boson and may deserve particular scrutiny in light of the CDF measurement.

Despite these caveats, non-vanishing oblique parameters in the global electroweak fit are widely interpreted as signals of physics beyond the Standard Model. In particular, the definition \eqref{rhodef} makes clear that $\rho$, and therefore $T$, should be expected to rise if the $W$ boson is heavier relative to the $Z$ than predicted in the Standard Model. Before the anomalous CDF measurement of $m_W$, the global fit --- with world average measured $W$ mass \cite{Haller:2018nnx,pdg2020}
\begin{align}
    m_W^{\text{PDG}} = 80.379 \pm 0.012\, \text{GeV}
\end{align}
used as the $m_W$ input --- stringently limited $S$, $T$, and $U$ both with and without the assumption of vanishing $U$, with best-fit values
\begin{align}
\nonumber S &= 0.04 \pm 0.11\ (0.04 \pm 0.08),\\
\nonumber T &= 0.09 \pm 0.14\ (0.08 \pm 0.07),\\
\text{and}\ \ \ U &= -0.02 \pm 0.11\ (\equiv 0)
\end{align}
compatible with $S=T=U=0$; \emph{i.e.}, the Standard Model, at the level of two standard deviations ($2\sigma$) \cite{Haller:2018nnx}.
\begin{figure}\label{STfit}
\begin{center}
    \includegraphics[scale=0.7]{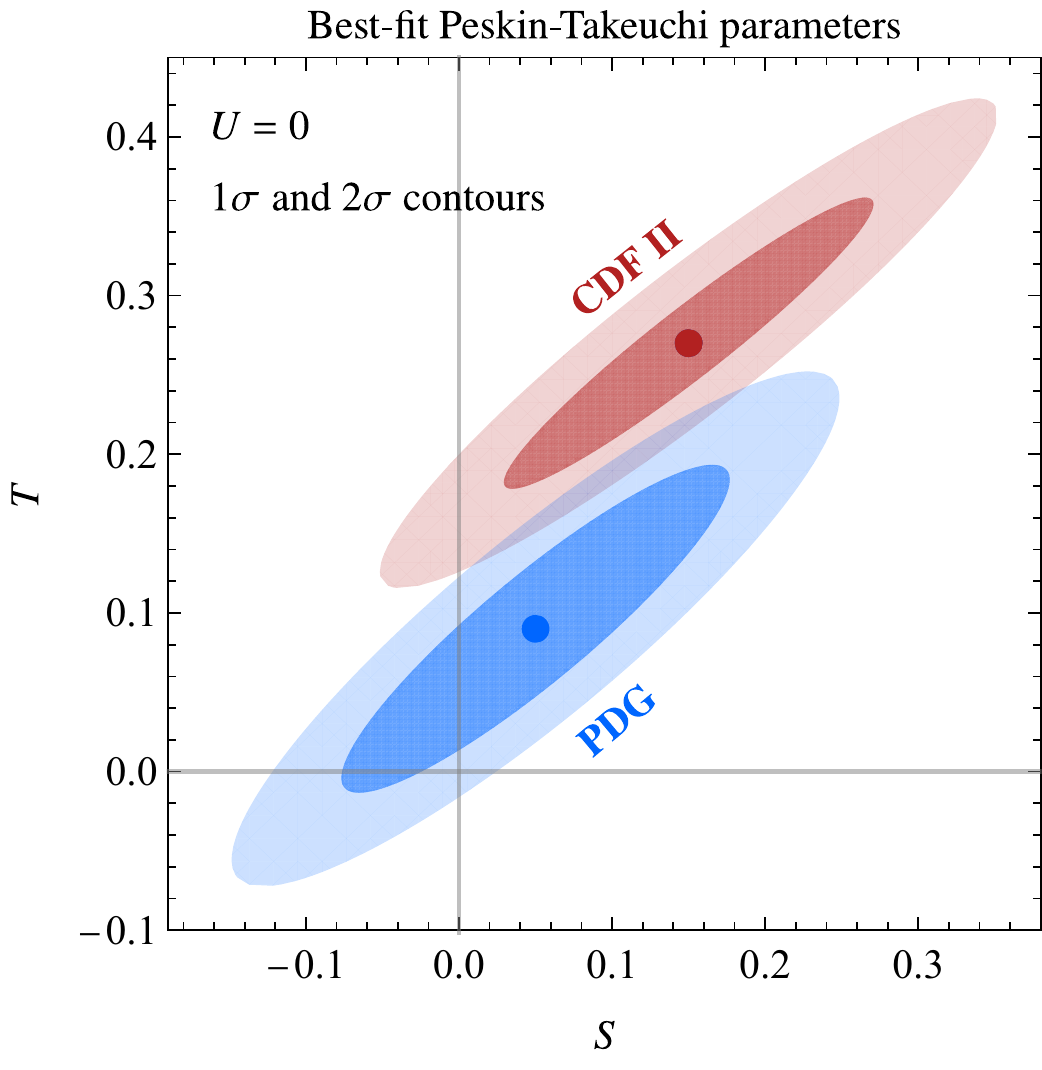}
\end{center}
\caption{Estimated \cite{Lu:2022bgw} best-fit values of oblique-correction parameters $S$ and $T$, assuming vanishing $U$, for (``PDG'') world average $W$ boson mass preceding CDF measurement and (``CDF II'') CDF result for $m_W$. Standard Model expectation, by definition, is at origin ($S=T=0$).}
\end{figure}The global fit has already been updated to use either $m_W^{\text{CDF II}}$ \eqref{CDFmeas}, or an uncorrelated average of the old and new values, by several groups following the strategy of the Gfitter group, using both \textsc{Gfitter} itself \cite{Lu:2022bgw} and other publicly available (\textsc{HEPFit}; \cite{deBlas:2022hdk}) and private \cite{Asadi:2022xiy} software producing similar results from a simplified global fit. We take as an estimate the results of \cite{Lu:2022bgw}, which finds best-fit values for $m_W^{\text{CDF II}}$ of
\begin{align}
    S = 0.15 \pm 0.08\ \ \ \text{and}\ \ \ T = 0.27 \pm 0.06
\end{align}
with $U=0$, as a benchmark to target with our models. These results are reproduced, along with a validating reproduction of the $S$-$T$ ($U=0$) fit for $m_W^{\text{SM}}$, in \hyperref[STfit]{Figure 1}. All $U=0$ global fits of which we are aware show only a small overlap between the $2\sigma$ preferred regions for $m_W^{\text{PDG/SM}}$ and $m_W^{\text{CDF II}}$, which reflects the significant tension introduced by the CDF measurement. The $U=0$ $m_W^{\text{CDF II}}$ fits still accommodate $S=0$ at the $2\sigma$ level, but significantly disfavor $T \leq 0$. It has already been noted, however, that allowing $U$ to float in the global fit eases the pressure on $S$ and $T$ at the cost of favoring large values of $\mathcal{O}(10^{-1})$ for $U$, which we mentioned above are difficult to produce \cite{Asadi:2022xiy}. We therefore emphasize again that the free-floating $U$ scenario should be considered in future work.

%% file: TeX/3_Model.tex
\section{Building models with sizable \emph{T}}
\label{s3}

If the $W$ boson is heavier than predicted within the Standard Model, but the SM $Z$ mass prediction is correct, then there is upward pressure on $T$ and perhaps $S$ if $U=0$. This reflects the fact that unequal changes to the weak boson masses violate the custodial $\mathrm{SU}(2)_V$ symmetry of the Standard Model. This observation suggests new electroweak physics. In this section, we describe two simple extensions of the Standard Model that can produce non-negative $S$ and $T$. We compute the oblique corrections in these models, explain how they arise, and discuss possible couplings to the Standard Model other than those guaranteed by gauge interactions.

\subsection{A model with two color-charged weak doublets}
\label{s3.1}

Extensions of the Standard Model featuring additional $\mathrm{SU}(2)_{\text{L}}$ doublets abound, both as independent schemes (\emph{viz}. the many two-Higgs-doublet models) and embedded in \emph{e.g.} supersymmetric models (MSSM and the like), and have enjoyed plenty of attention for many years owing to their rich and varied phenomenology. These additional doublets often share weak hypercharge with the SM Higgs doublet $(Y = 1/2)$, but the MSSM is a notable interesting case with oppositely charged novel doublets \cite{Martin:1997sp}. Either way, the electrically neutral scalars in these models attain vacuum expectation values (VEVs) and mix to produce the SM-like Higgs boson (with two other neutral scalars and two charged scalars). Some models in this large class have already been proposed to accommodate the CDF $m_W$ measurement \cite{Heo:2022dey,Bahl:2022xzi,Song:2022xts}. 

To inject some variety into this discussion, we first propose a framework in which the Standard Model is extended by two new weak doublets $\Phi_{\text{a}},\Phi_{\text{b}}$ whose components also carry color charge. The novel field content and quantum numbers of this \emph{double-color-doublet model} are displayed in \hyperref[DCDtab]{Table II}.\renewcommand\arraystretch{2}
\begin{table}\label{DCDtab}
    \centering
    \begin{tabular}{|c|| c |c|}
    \toprule
\hline 
\ Multiplet\ \ & \ Components\ \ & \ $\mathrm{SU}(3)_{\text{c}} \times \mathrm{SU}(2)_{\text{L}} \times \mathrm{U}(1)_Y$\ \ \\ 
\hline
\hline
$\Phi_{\text{a}}$ & $(\varphi^+_{\text{a}},\ \varphi^0_{\text{a}})$ & \multirow{2}{*}[0ex]{$(\textbf{r},\boldsymbol{2},\tfrac{1}{2})$} \\ 
\cline{0-1}
$\Phi_{\text{b}}$ & $(\varphi^+_{\text{b}},\ \varphi^0_{\text{b}})$ & \\
\hline 
\bottomrule
\end{tabular} 
    \caption{Novel field content in the double-color-doublet model. See text and \hyperref[freedom]{Table III} for discussion of color representations $\textbf{r}$.}
\end{table}
\renewcommand\arraystretch{1}For the moment, we leave unspecified the $\mathrm{SU}(3)_{\text{c}}$ color representation \textbf{r} shared by these new doublets. In principle, we have some freedom in this regard, though the oblique corrections in this model depend on \textbf{r} and we review in \hyperref[s3.3]{Section III.C} how many color-charged scalars can be added to the Standard Model while preserving asymptotic freedom. In any event, the unbroken $\mathrm{SU}(3)_{\text{c}}$ symmetry observed in Nature forbids the electrically neutral scalars in the double-color-doublet model from attaining VEVs and mixing with the SM Higgs. In the interest of minimality, we neglect all possible tree-level couplings between these scalars and the SM Higgs and consider the concrete example
\begin{multline}\label{DCDlag}
   \mathcal{L} \supset (D_{\mu}\Phi_{\text{a}})^{\dagger} D^{\mu}\Phi_{\text{a}} + (D_{\mu}\Phi_{\text{b}})^{\dagger} D^{\mu}\Phi_{\text{b}}\\ - m_{\text{a}}^2 \tr \Phi_{\text{a}}^{\dagger i} \Phi_{\text{a}i} - m_{\text{b}}^2 \tr \Phi^{\dagger i}_{\text{b}} \Phi_{\text{b}i} - (\Delta_{\text{ab}}^2 \tr \Phi_{\text{a}}^{\dagger i} \Phi_{\text{b}j} + \text{H.c.})\\ + \text{neglected terms permitted by symmetries}
\end{multline}
with $D^{\mu}$ the $\mathrm{SU}(3)_{\text{c}} \times \mathrm{SU}(2)_{\text{L}} \times \mathrm{U}(1)_Y$ covariant derivative and with traces ($\tr$) over implicit color indices. Crucially, mixing between color doublets is effected by the term proportional to $\Delta_{\text{ab}}$, which splits certain pairs of complex scalars but does not split the CP-even and -odd components of the electrically neutral complex scalars $\varphi_{\text{a,b}}^0$. In the end, the terms \eqref{DCDlag} produce a neutral-scalar mass matrix of the form
\begin{align}\label{DCDmass}
\mathcal{L} \supset -\begin{pmatrix}
\varphi_{\text{a}}^{0\dagger} & \varphi_{\text{b}}^{0\dagger} \end{pmatrix}
\begin{pmatrix}
m_{\text{a}}^2 & \Delta_{\text{ab}}^2\\
\Delta_{\text{ab}}^2 & m_{\text{b}}^2\end{pmatrix}
\begin{pmatrix}
\varphi_{\text{a}}^0\\
\varphi_{\text{b}}^0
\end{pmatrix},
\end{align}
whose eigenvalues are 
\begin{align}
    \begin{matrix}
    m_2^2\\[1ex]
    m_1^2
    \end{matrix}\ =\ \frac{1}{2}\left\lbrace m_{\text{a}}^2 + m_{\text{b}}^2 \pm \left[(m_{\text{a}}^2 - m_{\text{b}}^2)^2 + 4(\Delta_{\text{ab}}^2)^2\right]^{1/2}\right\rbrace
\end{align}
and which can be diagonalized by an orthogonal matrix of the form
\begin{align}
    \bt{O}_{\Phi} = \begin{pmatrix}
    \cos \theta_{\Phi} & \sin \theta_{\Phi}\\
    -\sin \theta_{\Phi} & \cos \theta_{\Phi}\end{pmatrix}
\end{align}
with mixing angle $\theta_{\Phi}$ satisfying
\begin{align}
\sin 2\theta_{\Phi} = \frac{2\Delta_{\text{ab}}^2}{m_2^2 - m_1^2}\ \text{and}\ \cos 2\theta_{\Phi} = \frac{m_{\text{a}}^2 - m_{\text{b}}^2}{m_2^2 - m_1^2}.
\end{align}
With this choice of mixing matrix, the neutral mass eigenstates $\Phi_{12}^0 = (\varphi_1^0,\ \varphi_2^0)^{\text{T}}$ are related to the weak eigenstates $\Phi_{\text{ab}}^0 = (\varphi_{\text{a}}^0,\ \varphi_{\text{b}}^0)^{\text{T}}$ according to $\Phi_{\text{ab}}^0 = \bt{O}_{\Phi}\Phi_{12}^0$. The charged states $\Phi_{12}^+ = (\varphi_1^+,\ \varphi_2^+)^{\text{T}}$ and $\Phi_{\text{ab}}^+ = (\varphi_{\text{a}}^+,\ \varphi_{\text{b}}^+)^{\text{T}}$ share an identical relationship, so that the physical spectrum is of the form
\begin{align}
\nonumber    \varphi_2^+, \varphi_2^0\ \ &\text{degenerate with mass}\ m_2\\
    \varphi_1^+, \varphi_1^0\ \ &\text{degenerate with mass}\ m_1.
\end{align}

The gauge interactions in the first line of \eqref{DCDlag} produce couplings at tree level between SM gauge bosons and both charged and neutral scalars, which in principle generate new contributions at one-loop order to the $W$, $Z$, and $\gamma$ self-energies and (therefore) $S$, $T$, and $U$. Representative diagrams for the $W$ boson self-energy are displayed as an example in \hyperref[Wmassfig]{Figure 2}.
\begin{figure*}\label{Wmassfig}
\begin{align*}
\scalebox{0.75}{\begin{tikzpicture}[baseline={([yshift=-0.8ex]current bounding box.center)},xshift=12cm]
\begin{feynman}[large]
\vertex (i1);
\vertex [right = 1.25cm of i1] (i2);
\vertex [right= 1.25cm of i2] (g1);
\vertex [right=1.25cm of g1] (g2);
\vertex [above right = 0.75cm and 1cm of i2] (p1);
\vertex [below right = 0.75cm and 1cm of i2] (p2);
\vertex [above right=1.5 cm of g1] (v1);
\vertex [below right=1.5cm of g1] (v2);
\diagram* {
(i1) -- [ultra thick, charged boson] (i2), %momentum={[arrow shorten=0.15]$p$}] (i2),
(i2) -- [ultra thick, charged scalar,  half left, looseness=1.7] (g1),
(g1) -- [ultra thick, scalar, half left, looseness=1.7] (i2),
(g1) -- [ultra thick, charged boson] (g2)
};
\end{feynman}
\node at (0.32,0.34) {$W^{\pm}$};
\node at (1.875,1.05) {$\varphi_I^+$};
\node at (1.875,-1) {$\varphi^0_J$};
\end{tikzpicture}}\ \ \ \ +\ \ \ \  \scalebox{0.75}{\begin{tikzpicture}[baseline={([yshift=-5.9ex]current bounding box.center)},xshift=12cm]
\begin{feynman}[large]
\vertex (i1);
\vertex [right = 1.75cm of i1] (i2);
\vertex [right= 1.75cm of i2] (g1);
\vertex [above=1.25cm of i2] (l1);
\vertex [above right = 0.75cm and 1cm of i2] (p1);
\vertex [below right = 0.75cm and 1cm of i2] (p2);
\vertex [above right=1.5 cm of g1] (v1);
\vertex [below right=1.5cm of g1] (v2);
\diagram* {
(i1) -- [ultra thick, charged boson] (i2), %momentum={[arrow shorten=0.15]$p$}] (i2),
(i2) -- [ultra thick, charged boson] (g1)
};
\end{feynman}
\draw[dashed, ultra thick, black] (1.75,0.64) circle[radius=0.64cm];
\node at (0.32,0.34) {$W^{\pm}$};
\node at (2.55,1.3) {$\varphi_I^+$};
\node[isosceles triangle,
    draw,
    fill=black,
    inner sep = 1.7pt] (T)at (1.75,1.28){};
\end{tikzpicture}}\ \ \ \ +\ \ \ \  \scalebox{0.75}{\begin{tikzpicture}[baseline={([yshift=-5.9ex]current bounding box.center)},xshift=12cm]
\begin{feynman}[large]
\vertex (i1);
\vertex [right = 1.75cm of i1] (i2);
\vertex [right= 1.75cm of i2] (g1);
\vertex [above=1.25cm of i2] (l1);
\vertex [above right = 0.75cm and 1cm of i2] (p1);
\vertex [below right = 0.75cm and 1cm of i2] (p2);
\vertex [above right=1.5 cm of g1] (v1);
\vertex [below right=1.5cm of g1] (v2);
\diagram* {
(i1) -- [ultra thick, charged boson] (i2), %momentum={[arrow shorten=0.15]$p$}] (i2),
(i2) -- [ultra thick, charged boson] (g1)
};
\end{feynman}
\draw[dashed, ultra thick, black] (1.75,0.64) circle[radius=0.64cm];
\node at (0.32,0.34) {$W^{\pm}$};
\node at (2.55,1.3) {$\varphi^0_I$};
\end{tikzpicture}}
\end{align*}
\caption{Diagrams contributing to the $W$ boson self-energy $\Pi_{WW}^{\text{new}}(p_W^2)$ in the double-color-doublet model with $\mathrm{SU}(2)_{\text{L}}$ doublets $\Phi_{\text{a}}^{\text{T}} = (\varphi_{\text{a}}^+,\ \varphi_{\text{a}}^0)$ and $\Phi_{\text{b}}^{\text{T}} = (\varphi_{\text{b}}^+,\ \varphi_{\text{b}}^0)$. Displayed are (complex) mass eigenstates $W^{\pm}$ and $\varphi^{\pm}_I, \varphi^0_I$, $I \in \{1,2\}$.}
\end{figure*}
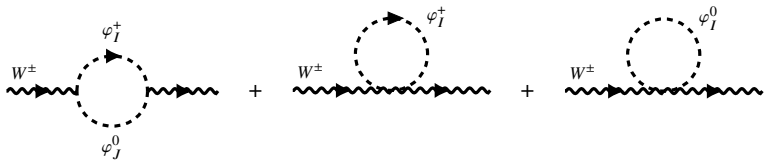These $W$ diagrams are instructive: the mixing term proportional to $\Delta_{\text{ab}}^2$ generates mixed loops containing \emph{e.g.} $\varphi^+_1, \varphi^0_2$ that do not identically cancel. Put another way, the mixing term induces splitting between charged and neutral scalars --- which is often generated uniquely by the Higgs VEV --- which results in unequal contributions to the electrically charged and neutral components of the $\mathrm{SU}(2)_{\text{L}}$ triplet of $W^A$ bosons. This effect is exactly what is measured by the $T$ parameter, so these diagrams are telling us to expect non-vanishing $T$. Altogether, the relevant double-color-doublet oblique corrections are given by
\begin{align}\label{DCDst}
S = 0\ \ \ \text{and}\ \ \ \delta\rho = \alpha T = \frac{\alpha}{64\pi}\frac{r}{m_W^2} \left(\frac{\sin 2 \theta_{\Phi}}{ \sin \theta_{\text{w}}}\right)^2 f(m_1^2,m_2^2),
\end{align}
where $r$ is the dimension of the doublets' $\mathrm{SU}(3)_{\text{c}}$ representation \textbf{r} (\emph{i.e.} the number of colors), and where
\begin{align}
f(m_1^2,m_2^2) = m_1^2 + m_2^2 - \frac{2 m_1^2 m_2^2}{m_1^2-m_2^2} \ln \frac{m_1^2}{m_2^2}
\end{align}
is a function endemic to models with novel scalar contributions to gauge boson self-energies \cite{Lavoura:1993nq,Manohar:2006ga}. Strikingly, $S$ vanishes for all masses and mixings, but $T$ is proportional to the extent of mixing, as measured by $\sin 2 \theta_{\Phi}$, and it can be made quite large. We have checked for consistency that $U$ is finite, but it turns out not to vanish identically and cannot be expressed particularly compactly. It is, however, uniformly smaller than $T$ --- in keeping with the power counting discussed in \hyperref[s2]{Section II} --- and we neglect it going forward.

\subsection{A weak + strong biadjoint model}
\label{s3.2}

We also consider a model with a field $\Sigma$ transforming in the adjoint representations of both $\mathrm{SU}(2)_{\text{L}}$ and $\mathrm{SU}(3)_{\text{c}}$. The notation and quantum numbers for this multiplet are given in \hyperref[TRIPtab]{Table III}.\renewcommand\arraystretch{2}
\begin{table}\label{TRIPtab}
    \centering
    \begin{tabular}{|c|| c |c|}
    \toprule
\hline 
\ Multiplet\ \ & \ Components\ \ & \ $\mathrm{SU}(3)_{\text{c}} \times \mathrm{SU}(2)_{\text{L}} \times \mathrm{U}(1)_Y$\ \ \\ 
\hline
\hline
$\Sigma$ &\ $(\Sigma^+,\ \Sigma^0,\ \Sigma^-)$\ \ & $(\textbf{r},\boldsymbol{3},0)$\\
\hline 
\bottomrule
\end{tabular} 
    \caption{Novel field content in the biadjoint model.}
\end{table}
\renewcommand\arraystretch{1}Here we express the field (suppressing for a moment its $\mathrm{SU}(3)_{\text{c}}$ index) as
\begin{multline}
\Sigma = \bt{t}_{\boldsymbol{2}}^A \Sigma^A = \frac{1}{2}\begin{pmatrix}
\Sigma^0 & \sqrt{2}\Sigma^+ \\
\sqrt{2}\Sigma^- & -\Sigma^0
\end{pmatrix},\\ \text{with}\ \ \ \Sigma^+ = \frac{\Sigma^1 -\ii \Sigma^2}{\sqrt{2}}, \ \ \Sigma^- = \frac{\Sigma^1 +\ii \Sigma^2}{\sqrt{2}},
\end{multline}
where $\bt{t}_{\boldsymbol{2}}^A = \sigma^A/2$ ($A \in \{1,2,3\}$) are the generators of the $\boldsymbol{2}$ of $\mathrm{SU}(2)$. Its potential can be minimally written as
\begin{multline}
    V(H,\Sigma) \supset \lambda  \left(H^{\dagger i} H_i -  \frac{1}{2}\,v^2\right)^2\\ + m^2_\Sigma\tr \Sigma^\dagger \Sigma+ \lambda_\Sigma \tr\,(\Sigma^\dagger \Sigma)^2 + A_\Sigma\, f_{abc}\epsilon^{ABC}\,\Sigma_A^a\Sigma_B^b\Sigma_C^c.
    \end{multline}
The trilinear $A_\Sigma$ term is gauge invariant, but we shall assume it is small enough to safely neglect. Weak triplets and their contributions to precision observables have been well studied \cite{Blank:1997qa,Khandker:2012zu}. If the triplet has no other nonzero quantum numbers, the neutral component will generically attain a nonzero vacuum expectation value, which contributes to $T$ at tree level. In the model studied in this work, the triplet is charged under color. Therefore, there is no VEV, and contributions to electroweak variables occur at the one loop level. At tree level, the terms in the above potential do not produce a mass splitting between the neutral and charged components of the biadjoint. One way to generate a mass splitting is to add an extra color-octet field $O$ in the $(\boldsymbol{8},\boldsymbol{1},0)$ representation of the SM gauge group. With this additional field, we may add the terms 
\begin{equation}
    V(H,O) \supset m_O^2|O|^2 + \lambda_{O\Sigma}\, O^a H^\dagger \Sigma^a H 
\end{equation}
to the potential, and after Higgs VEV insertions, the electrically neutral part $\Sigma^0$ of the biadjoint mixes with the weak singlet $O$. This pulls the mass of this component away from the common value $m_\Sigma$. With this idea as a proof of concept, we take the charged and neutral mass eigenstates to be non-degenerate, and we denote them by $m_+$ and $m_0$ in the discussion below.

General expressions for $S$ and $T$ parameters from weak multiplets may be found in \cite{Lavoura:1993nq}. In this model, $S$ reduces to
\begin{equation}
    S=\frac{4N_c c_\text{w}^4}{\pi} \left[-\frac{4}{9} + \frac{4}{3}\frac{m_+^2}{m_Z^2}+\frac{1}{12}\, \Delta\!\left(\frac{m_+}{m_Z}\right)^{3/2} F\!\left(\frac{m_+}{m_Z}\right) \right],
\end{equation}
where
\begin{multline}
    \Delta(x) = 4x^2-1\\ \text{and}\ \ \ F(x) = -2\left[\arctan\frac{1}{\sqrt{\Delta(x)}}-\arctan\frac{-1}{\sqrt{\Delta(x)}} \right]. %\ln\frac{2x-1+\beta(x)}{2x-1-\beta(x)}.
\end{multline}
$S$ is in general nonzero, and it depends only on the mass of the charged state. However, for $m_+\gg m_Z$, the contribution to $S$ tends to 0, as one might expect from the decoupling limit. The contribution to the $T$ parameter is very similar in form to the case of the double-color-doublet model:
\begin{equation}
    T=\frac{N_c}{8\pi} \frac{1}{(s_{\text{w}} c_{\text{w}})^2}\left[\frac{m_+^2+m_0^2}{m_Z^2} -\frac{2m_0^2 m_+^2 \ln(m_+^2/m_0^2)}{m_Z^2 (m_+^2-m_0^2)}\right].
\end{equation}
$T$ only depends on the splitting between the masses, so even heavy states may have a sizeable impact.

\subsection{Embedding in the Standard Model}
\label{s3.3}

Before we investigate whether either of these models are viable in light of the CDF measurement of $m_W$, we pause to discuss two basic model-building considerations that apply to some extent to both of our scenarios. The first is a concern common to any model with additional color-charged states. In particular, it is well known that scalars enjoying couplings to gluons contribute to the one-loop $\mathrm{SU}(3)_{\text{c}}$ $\beta$ function according to \cite{PSQFT}
\begin{align}\label{betafn}
    \beta(g_3) - \beta_{\text{SM}}(g_3) = \frac{g_3^3}{12\pi}\,N_{\varphi} T_{\textbf{r}_{\varphi}}\ \ \ \text{with}\ \ \ \beta_{\text{SM}}(g_3) = -\frac{7g_3^3}{4\pi},
\end{align}
where $N_{\varphi}$ is the number of complex scalar fields transforming in representation $\textbf{r}$ of $\mathrm{SU}(3)_{\text{c}}$ and $T_{\textbf{r}_{\varphi}}$ is the Dynkin index, or generator normalization, of $\textbf{r}$:
\begin{align}
    \tr \bt{t}_{\textbf{r}}^a \bt{t}_{\textbf{r}}^b \equiv T_{\textbf{r}}\delta^{ab}.
\end{align}
Given that a formidable array of measurements of the strong coupling $g_3$ have found excellent agreement with the SM prediction of $\beta(g_3) < 0$ (\emph{asymptotic freedom}) \cite{BETHKE2007351}, it is reasonable to limit the number of additional color-charged scalar fields in extensions of the Standard Model to avoid changing the sign of $\beta(g_3)$. We use \eqref{betafn} as a rough leading-order guide to this end. If we supply the Dynkin indices $T_{\textbf{r}}$, we can compute the maximum number of allowed (complex) scalars in representations of $\mathrm{SU}(3)_{\text{c}}$ accessible at hadron colliders (\emph{i.e.} up to $\boldsymbol{8} \otimes \boldsymbol{8} \supset \boldsymbol{27}$), which is a function of the number of weak doublets. Upper bounds for such representations (assuming no other new color-charged fields) are displayed in \hyperref[freedom]{Table IV}.\renewcommand\arraystretch{2}
\begin{table}\label{freedom}
    \centering
    \begin{tabular}{|c|| c |c|}
    \toprule
\hline 
\ \textbf{r} of $\mathrm{SU}(3)_{\text{c}}$\ \ & \ $T_{\textbf{r}}$\ \ & \ $N_{\varphi}^{\text{max}}$\ \ \\ 
\hline
\hline
$\boldsymbol{3}$ & $\tfrac{1}{2}$ & 42 \\ 
\hline 
$\boldsymbol{6}$ & $\tfrac{5}{2}$ & 8\\
\hline
$\boldsymbol{8}$ & 3 & 7\\
\midrule
$\boldsymbol{10}$ & $\tfrac{15}{2}$ & 2\\
\hline
$\boldsymbol{15}$ & 10 & 2\\
\hline
$\boldsymbol{24}$ & 25 & 0\\
\hline 
\bottomrule
\end{tabular} 
    \caption{Number of additional complex color-charged scalars that can be added to QCD without compromising asymptotic freedom at one-loop order in the absence of other new fields.}
\end{table}
\renewcommand\arraystretch{1}We find that only scalars in the fifteen-dimensional representation(s) and smaller can be accommodated at leading order, with a single $\boldsymbol{24}$ already reversing the sign of $\beta(g_3)$. We also find that only fairly small numbers of weak multiplets are allowed for smaller representations: for example, a model with complex color-sextet ($\boldsymbol{6}$) scalars can only fit four $\mathrm{SU}(2)_{\text{L}}$ doublets, and a model with color-octet scalars can only have two. Nevertheless, there are enough viable configurations with respect to this criterion for both frameworks introduced in this section for us to proceed.

The other important question is far more open ended and concerns the full suite of couplings between our novel scalars and the Standard Model, which are not in general limited to gauge interactions. While we delay a comprehensive discussion to future work, we note here that only color-octet multiplets enjoy non-minimal couplings to SM gauge bosons that are (a) independent of quarks and (b) parametrically large; \emph{i.e.}, of small order in an effective field theory expansion. In the case of the double-color-doublet model, we have
\begin{multline}\label{DCDeff}
    \mathcal{L}_{\Phi}^{\text{eff}} \supset \frac{\kappa_{\Phi gg}}{\Lambda_{\Phi gg}^2}\,d_{abc}\, (H^{\dagger} \Phi_{\text{a,b}}^a)\, G^{b}_{\mu \nu}G^{c\,\mu \nu}\\ + \frac{\kappa_{\Phi gB}}{\Lambda_{\Phi gB}^2}\, (H^{\dagger} \Phi_{\text{a,b}}^a)\, B_{\mu\nu} G_{a}^{\mu \nu}\\ + \frac{\kappa_{\Phi gW}}{\Lambda_{\Phi gW}^2}\, (H^{\dagger} \bt{t}_{\boldsymbol{2}}^A \Phi_{\text{a,b}}^a)\, W_{A\,\mu\nu}G_{a}^{\mu \nu}.
\end{multline}
As elsewhere, $\mathrm{SU}(2)_{\text{L}}$ fundamental indices are contracted within parentheses. We have made explicit the generators $\bt{t}_{\boldsymbol{2}}^A$, $A \in \{1,2,3\}$, of the $\boldsymbol{2}$ of $\mathrm{SU}(2)$. These operators couple electrically neutral color-octet scalars to gluon-gluon, gluon-$Z$ and gluon-photon pairs at effective dimension five after insertion of a SM Higgs VEV. Similar operators can be written for color doublets with opposite weak hypercharge after reconfiguring the Higgs doublets \cite{Carpenter:2021gpl}. The biadjoint model faces a similar situation if the new scalars are sequestered from quarks. Explicitly, we can write
\begin{multline}\label{TReff}
\mathcal{L}_{\Sigma}^{\text{eff}} \supset \frac{\kappa_{\Sigma gW}}{\Lambda_{\Sigma gW}}\, \Sigma _A^a W^A_{\mu\nu} G_a^{\mu \nu}\\ + \frac{\kappa_{\Sigma gB}}{\Lambda_{\Sigma gB}^3}\, (H^{\dagger} \bt{t}_{\boldsymbol{2}}^A \Sigma _A^a H)\, B_{\mu\nu}G_{a}^{\mu\nu}\\ + \frac{\kappa_{\Sigma gg}}{\Lambda_{\Sigma gg}^3}\, d_{abc}\, (H^{\dagger} \bt{t}_{\boldsymbol{2}}^A \Sigma _A^a H)\, G^b_{\mu \nu} G^{c\,\mu \nu}
\end{multline}
The first operator has mass dimension five, and the others generate couplings of effective dimension five upon the insertion of Higgs VEVs. Decay channels for the electrically neutral component of $\Sigma $ in this scenario include $g\gamma$ and $gZ$, while the charged states decay to $gW^{\pm}$.

%% file: TeX/4_Numbers.tex
\section{Finding the right parameter space}
\label{s4}

Here we use the analytic results presented in \hyperref[s3]{Section III} to locate parameter space in our models capable of fitting the values of $S$ and $T$ favored by the CDF measurement of the $W$ boson mass as displayed in \hyperref[STfit]{Figure 1}. We also offer some thoughts about constraints and possible future searches for our novel scalars at the Large Hadron Collider.

To facilitate our analytical and numerical investigation, we have implemented our models in version 2.3.43 of \textsc{FeynRules} \cite{FR_OG,FR_2}, a model-building package loaded in \textsc{Mathematica}$^{\copyright}$ version 12.0 \cite{Mathematica}. We have directed \textsc{FeynRules} to produce input for the \textsc{Mathematica} package \textsc{FeynArts} version 3.11 \cite{FA}, which we have used in turn to draw the gauge boson self-energy diagrams required to compute the oblique corrections. We have evaluated the resulting amplitudes using \textsc{FeynHelpers} version 1.3.0 \cite{feynhelpers}, which conveniently integrates \textsc{FeynCalc} version 9.3.0 \cite{MERTIG1991345,FC_9.0,FC_9.3.0} with \textsc{Package-X} version 2.1.1 \cite{Patel:2017px}.

\subsection{Fitting \emph{S} and \emph{T} for a heavier \emph{W}}
\label{s4.1}

\begin{figure*}\label{DCDspace}
    \centering
    \includegraphics[scale=0.7]{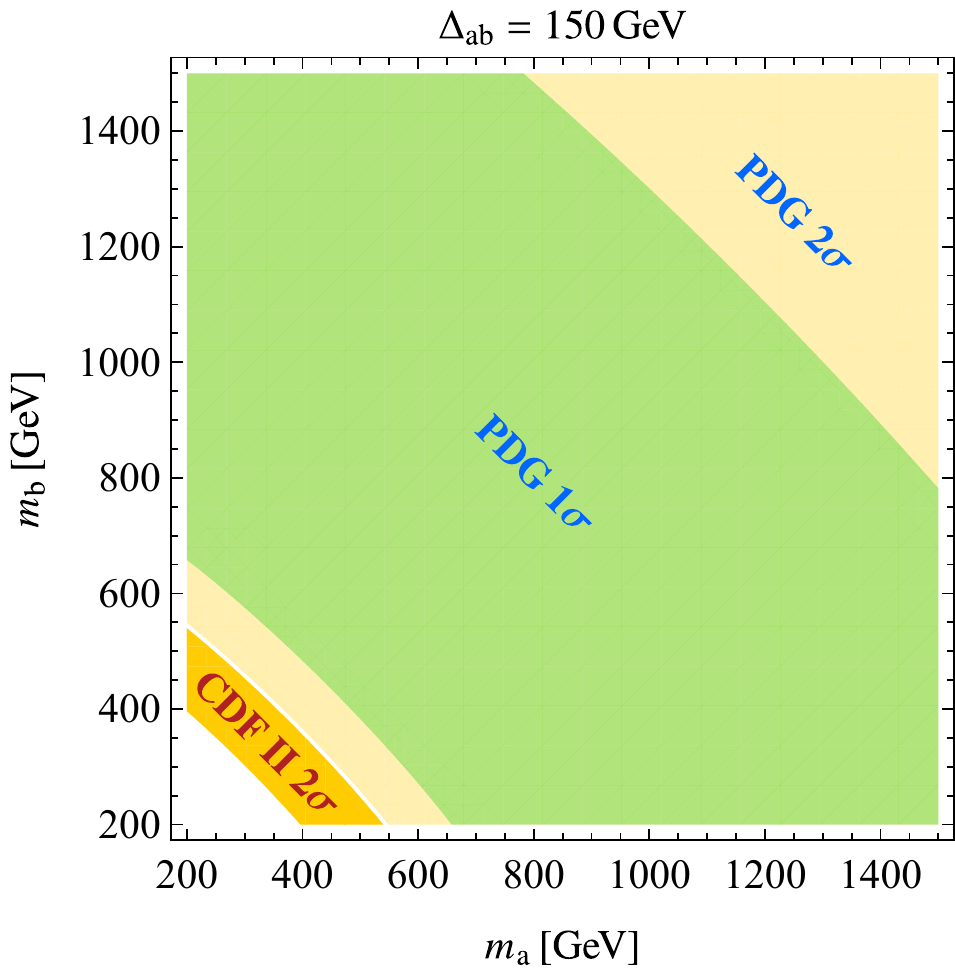}\hspace{1cm}\includegraphics[scale=0.7]{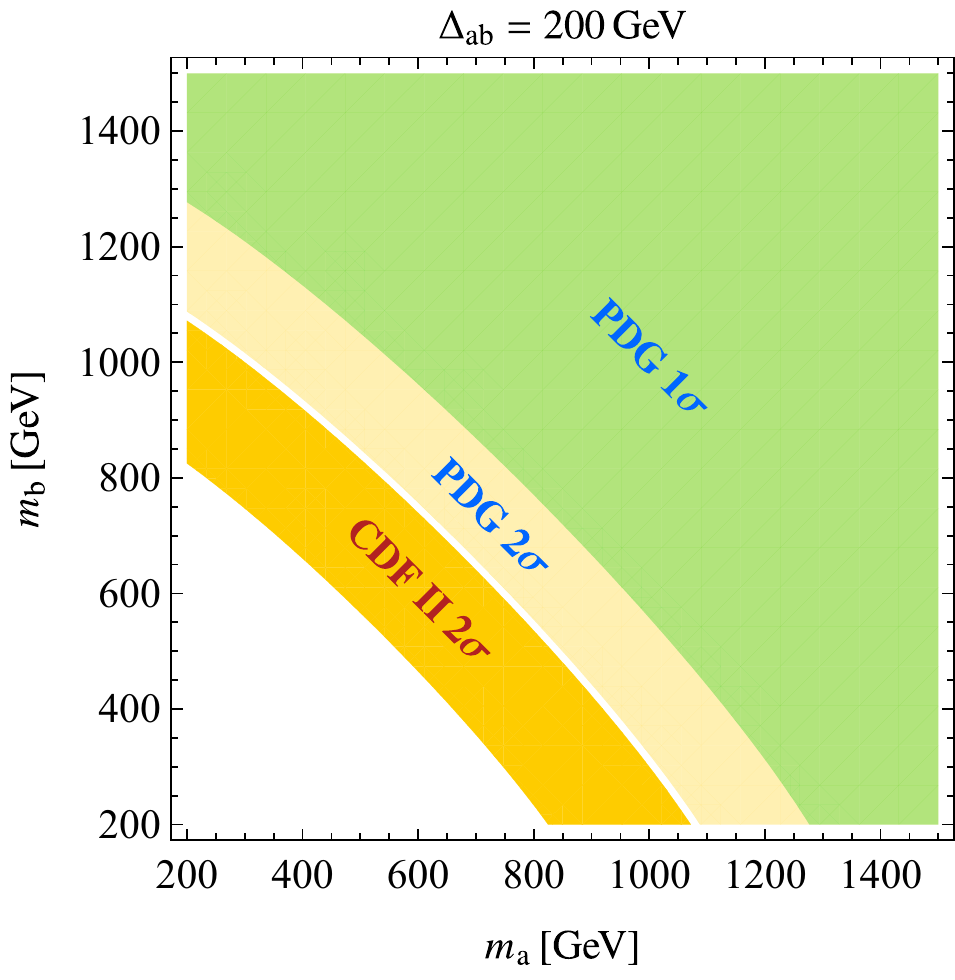}\\[4ex]\includegraphics[scale=0.7]{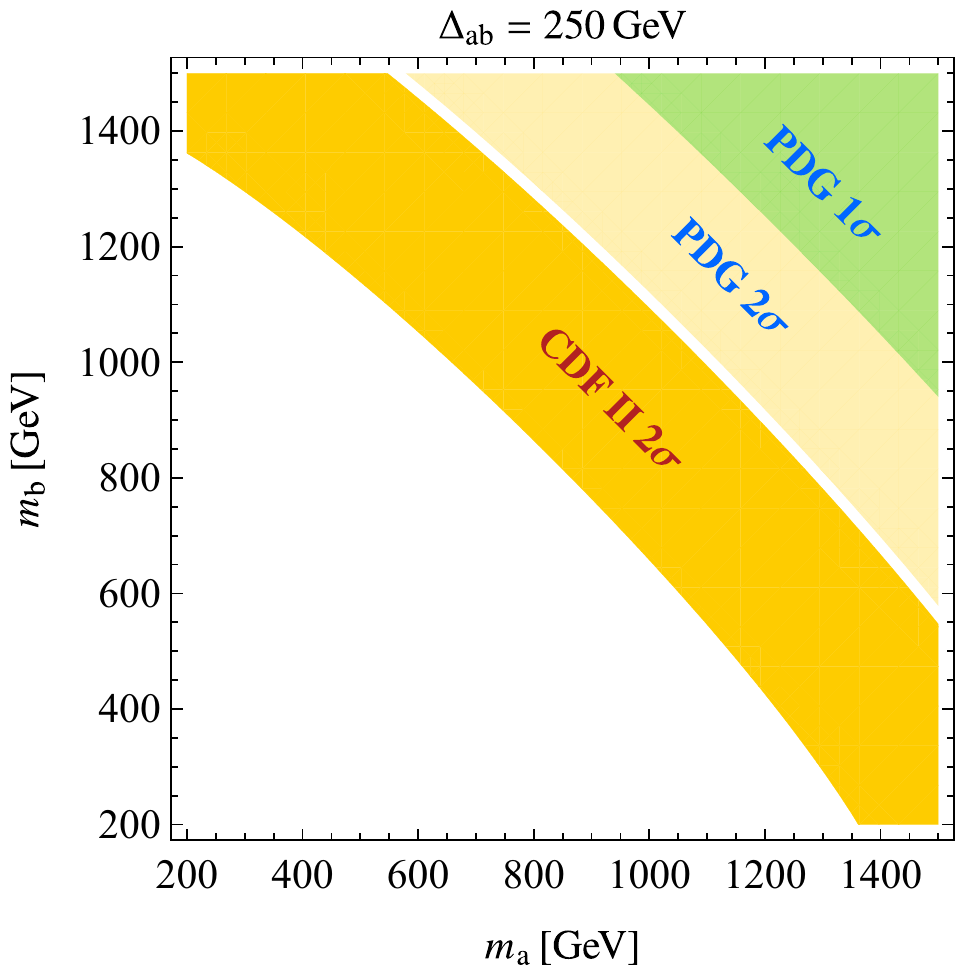}\hspace{1cm}\includegraphics[scale=0.7]{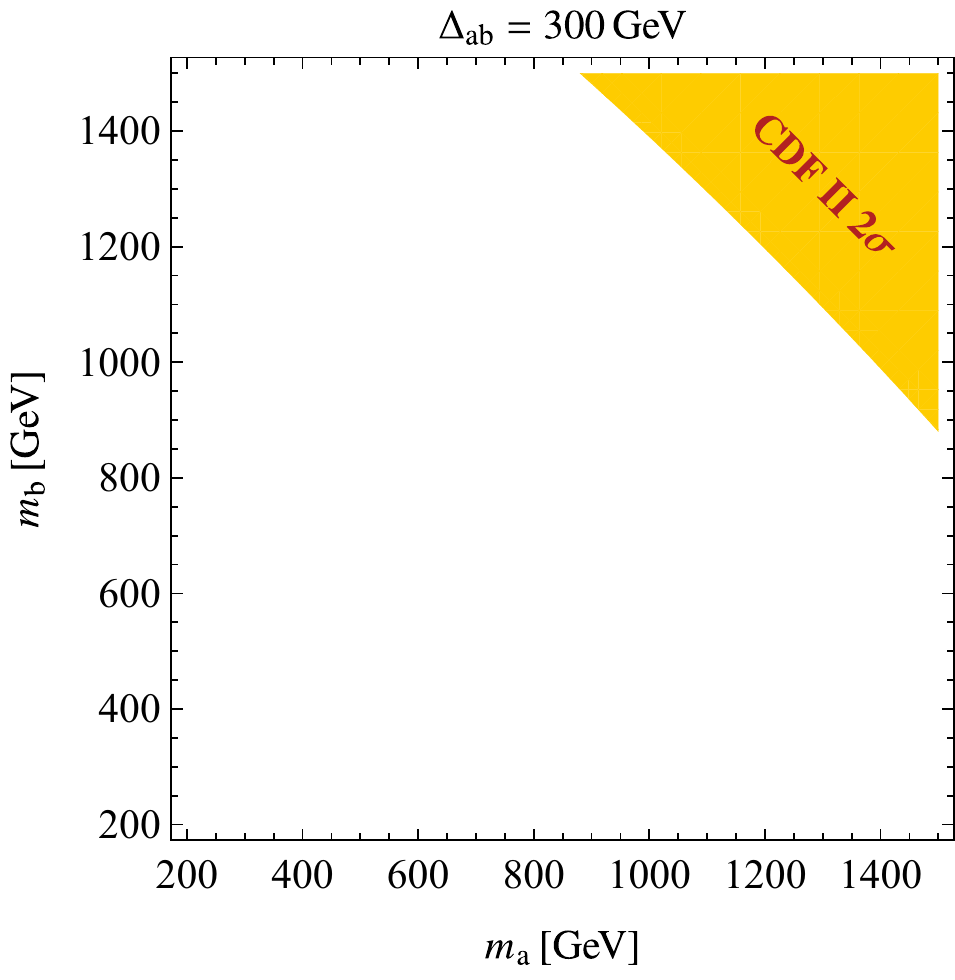}
    \caption{Parameter space in the double-color-doublet model with $r=8$ producing $T$ compatible with either the PDG or CDF global electroweak fits as estimated in \hyperref[STfit]{Figure 1}. Frames are arranged left to right, then top to bottom, with increasing mixing parameter $\Delta_{\text{ab}}$.}
\end{figure*}

\begin{figure}
    \centering
    \includegraphics[scale=0.7]{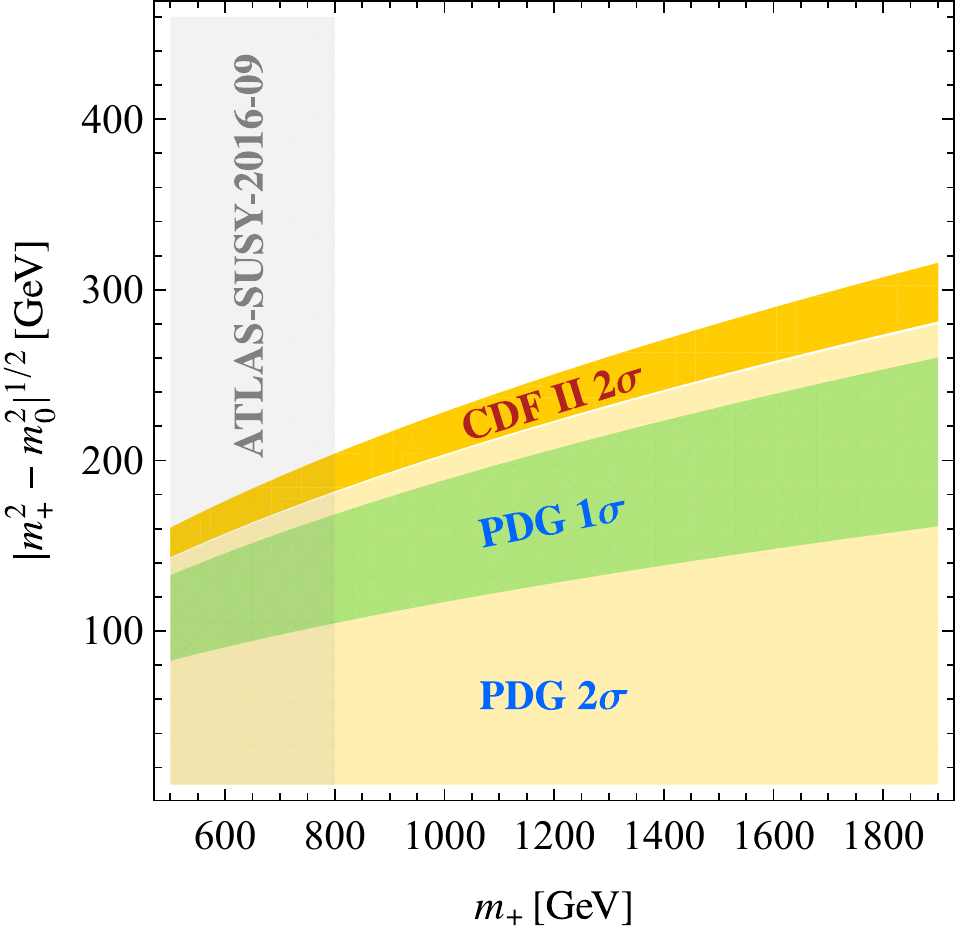}
    \caption{Space in the $(m_+,|m_0^2-m_+^2|^{1/2})$ plane of the bidoublet model where the $S$ and $T$ parameters are compatible with either the PDG or CDF global electroweak fits. Gray region estimates existing constraints from multijet searches (\emph{e.g.}, ATLAS-SUSY-2016-09 \cite{ATLAS:2017jnp}) on color-octet scalars decaying to gluons, as discussed in \cite{Carpenter:2020hyz}.}
    \label{fig:bimass}
\end{figure}

In accordance with the order of presentation in \hyperref[s3]{Section III}, we begin with the double-color-doublet model. Since this model contains four color-charged complex scalars, the largest $\mathrm{SU}(3)_{\text{c}}$ representation supported by this model without endangering asymptotic freedom at one-loop order (\emph{viz}. \hyperref[s3.3]{Section III.C}) is the adjoint with color factor $r=8$, in which the $T$ parameter \eqref{DCDst} is linear. For definiteness, we specialize to this color-octet scenario, which can be viewed as a generalization of the Manohar-Wise model \cite{Manohar:2006ga} with a second color-octet weak doublet. We display in \hyperref[DCDspace]{Figure 3} the regions of fundamental model parameter space, $(m_{\text{a}},m_{\text{b}},\Delta_{\text{ab}})$, capable of accommodating $S$ and $T$ for the old and new $W$ boson masses.

This figure consists of four panels to provide a (discrete) scan over three-dimensional parameter space. To interpret these panels, recall from \eqref{DCDst} that in the double-color-doublet model $S=0$ identically. \hyperref[STfit]{Figure 1} indicates that this model can therefore occupy the $2\sigma$ region preferred by the CDF value of $m_W$, but not the $1\sigma$ region. On the other hand, it can easily land in the $1\sigma$ and $2\sigma$ regions favored by the PDG average. Larger positive $T$ does not favor a particular hierarchy of $\{m_{\text{a}},m_{\text{b}}\}$, but it does prefer small mixing; \emph{i.e.}, $\Delta_{\text{ab}}$ small compared to $m_{\text{a}}$ and $m_{\text{b}}$. This is evident in the migration of the CDF $2\sigma$ preferred band toward heavier scalar masses with growing $\Delta_{\text{ab}}$. Notably, there is little to no overlap between the CDF and PDG $2\sigma$ preferred regions for $T$ at $S=0$, so while it is apparently easy to tune the double-color-doublet model to fit (at least the $2\sigma$ regions for) either model, there is no choice of inputs that accommodates both values of $m_W$ considered in this analysis.

Next, we perform a similar investigation of the biadjoint model. Figure \ref{fig:bimass} again shows model parameter space favored by the CDF and PDG global electroweak fits. The splitting between electrically charged and neutral scalars is presented in terms of the squared mass difference, which is most useful in translating to the fundamental parameter space of the model. We find that $T$ becomes dominant over $S$ when the scalar's mass rises above the weak scale, and that fairly small splittings between mass eigenvalues are favored by the CDF fit, with even smaller splittings and even degeneracy permitted by the PDG fit. As for the double-color-doublet model, there is no overlap between $2\sigma$ preferred regions owing to the nearly vanishing values of $S$ produced in this parameter space. \hyperref[fig:bimass]{Figure 4} finally displays a region with $m_+ < 800\,\text{GeV}$ --- which is the characteristic scalar mass in this model provided that the splitting is small --- that represents a conservative bound from Run 2 multijet searches (\cite{ATLAS:2017jnp} and similar) on pair-produced color-octet scalars decaying to gluons. We elaborate on collider constraints on our models immediately below.

\subsection{Confronting collider constraints}
\label{s4.2}

Before we conclude, we briefly consider collider constraints on the novel states in the models we have proposed to fit the CDF measurement of $m_W$. Since in both frameworks we specialized to scalars in the adjoint representation of $\mathrm{SU}(3)_{\text{c}}$, the operative question is how to constrain exotic color-octet scalars with $\mathrm{SU}(2)_{\text{L}} \times \mathrm{U}(1)_Y$ quantum numbers. Color octets have been studied at length in many different contexts \cite{Plehn:2008ae,Choi:2008ub,Carpenter:2020hyz,Carpenter:2020evo,Carpenter:2021vga}, including in at least one scheme like ours where they are arranged in a weak multiplet \cite{Manohar:2006ga}. Depending on the scenario, the most stringent constraints on color-octet scalars can come from dijet-resonance searches at the Large Hadron Collider, where a single electrically neutral scalar may be produced via gluon fusion through an operator of (perhaps effective) dimension five. (The first operator in \eqref{DCDeff} and the last operator in \eqref{TReff} play this role in our models after electroweak symmetry breaking.) But whether a large dijet signature is produced by the neutral scalar depends on its decays. If, for instance, the neutral scalar has an appreciable Yukawa-type coupling to quarks, which in the double-color-doublet model could arise from dimension-four operators of the form
\begin{align}
    \mathcal{L}_{\Phi} \supset -y_{\Phi}\, (\bar{Q}_{\text{L}}^i \Phi^a_{\text{a,b}})\, [\bt{t}_{\boldsymbol{3}}^a]_i^{\ j} d_{\text{R}j} + \text{H.c.} + \text{terms for $u_{\text{R}}$}
\end{align}
(with $\bt{t}_{\boldsymbol{3}}^a$ the generators of the $\boldsymbol{3}$ of $\mathrm{SU}(3)$), then its decays will be dominated by $\varphi^0 \to q\bar{q}$ and it may be severely constrained at LHC \cite{CMS:2018mgb}. But if the color doublets do not couple to quarks, then the signatures are dictated by the effective operators \eqref{DCDeff} and \eqref{TReff}, which permit decays to SM gauge bosons. The collider signatures in this case are highly sensitive to the hierarchy of masses: heavier neutral scalars decay to charged scalars and a $W$ boson and vice versa, with each scheme manifesting differently in a detector. In the double-color-doublet model with color-octet scalars, we find an unusual spectrum favoring the decays
\begin{align}
\nonumber    \varphi^+_2 &\to \varphi^0_1 + W^+\\[1ex]
             \varphi^0_2 &\to \varphi^+_1 + W^-
\end{align}
in the absence of tree-level couplings to the SM Higgs doublet (which are permitted by gauge symmetry, and should be fully appreciated in future work, but must be quite small to preserve the SM-like Higgs production cross section and branching fractions to gauge bosons \cite{Manohar:2006gz}). Scenarios with lighter electrically charged color-octet scalars are fascinating because, even if the corresponding neutral scalar is produced by gluon fusion with an appreciable cross section, the decay chain precludes a large dijet signature.

Electrically charged color-octet scalars, by contrast, cannot be singly produced: they must be pair produced, but they share with the neutral scalars a large pair-production cross section guaranteed by gauge interactions. Any search for pair-produced scalars in our models must confront the same questions about decay chains discussed just above: couplings to quarks often allow the LHC Run 2 dataset to exclude color-octet scalars at or just below the TeV scale \cite{Hayreter:2017wra,Carpenter:2020hyz}, but scalars decaying to other scalars and/or gauge bosons face weaker constraints \cite{Cacciapaglia:2020vyf}, with pair-produced scalars decaying solely to gluons excluded below approximately $800\,\text{GeV}$ \cite{Carpenter:2020hyz} as depicted in \hyperref[fig:bimass]{Figure 4}. This bound could be diluted for scalars also decaying to photons. It should be noted, however, that electrically neutral and charged color-octet scalars permitted to couple to quarks could be singly produced in association with said quarks and then decay to produce four-jet signatures. These production channels have been analyzed in the context of the Manohar-Wise model and found to have significant cross sections of $\mathcal{O}(10^2)\,\text{fb}$ for TeV-scale scalars at LHC with $\sqrt{s}=13\,\text{TeV}$, ultimately pushing limits to as high as $1.1\,\text{TeV}$ depending on benchmark \cite{Miralles:2019uzg,Eberhardt:2021ebh}. 

It is worth elaborating, before we conclude, on the possible role of the Standard Model(-like) Higgs in constraining these new color-charged weak multiplets. In the event that the SM Higgs doublet couples to a pair of new multiplets, it is possible to generate a trilinear coupling between two color-charged scalars and the Higgs boson upon insertion of a Higgs VEV. Such couplings can generate new diagrams at one-loop order that can dangerously enhance the cross section of Higgs boson production via gluon fusion. The same mechanism can also upset the branching fractions of the SM Higgs to photons and $Z$ bosons, which are measured to the 5\% level \cite{pdg2020} and therefore stringently constrain new physics in this sector. But by the same token, the Higgs VEV can be used to enhance the splitting between electrically charged and neutral scalars within the new weak multiplets and may be a valuable tool to produce larger oblique corrections. Realizations of our framework with sizable couplings to the SM Higgs doublet might therefore require a detailed balance in parameter space between splitting charged and neutral states, as desired, and preserving a single-Higgs production cross section compatible with experiment. Such scenarios may favor heavier scalars than considered in \hyperref[s4.1]{Section IV.A}, which should steadily decouple from the Higgs processes visible at the LHC. For example, in the biadjoint model, the splitting of around $25\,\text{GeV}$ needed to accommodate $m_W^{\text{CDF II}}$ can be produced by $\lambda_{O\Sigma} < 0.02$. Couplings of this size should produce acceptably small corrections to the Wilson coefficient of the effective Higgs-gluon-gluon coupling, well below the percent level for $m_O > 800\,\text{GeV}$ \cite{Boughezal:2010ry}.

%% file: TeX/5_Conclusions.tex
\section{Conclusions}
\label{s5}

The recently announced measurement by the CDF Collaboration of a $W$ boson mass significantly heavier than the Standard Model expectation demands explanation --- a demand that has already been satisfied in droves in a very short period of time. One emerging avenue of study is directed at supplying larger (positive) oblique corrections, parametrized by the Peskin-Takeuchi parameters $S,\,T,\,U$, in accordance with a global electroweak fit revised to accommodate the CDF measurement. As part of this effort, we have explored in this work some non-standard scenarios with weak scalar doublets whose neutral components do not attain vacuum expectation values (VEVs). We have demonstrated that such models are capable of generating, in particular, comparatively large $\mathcal{O}(10^{-1})$ values of $T$ --- with the aid of (a) VEVless mechanisms for breaking degeneracy between electrically charged and neutral scalars and (b) appreciably (though not arbitrarily) large color factors. We have provided analytic expressions for at least $S$ and $T$ in these models and quantitatively explored their ability to support a heavier $W$ boson. We have also considered the couplings of color-charged weak doublets to the Standard Model in a model-independent fashion and discussed possible collider constraints on the scalars in our models.

The need for new models to explain experimental anomalies, particularly in the electroweak sector, is far from new and has been intensifying for some time. There is a well known and increasingly perplexing constellation of muon anomalies, which have generated their own corpus of theoretical proposals --- some involving the $W$ mass --- and the CDF $m_W$ measurement is not even the first electroweak precision observable to threaten the Standard Model, with LEP showing a $2.8\sigma$ discrepancy in the $b$-quark forward-backward asymmetry $A_{\text{FB}}^{0,b}$ by 2006 \cite{ALEPH:2005ab}. We would be remiss if we did not also note the $6\sigma$ excess in electron-like events produced jointly by the Liquid Scintillator Neutrino Detector and MiniBooNE experiments \cite{miniboone_2018}. Nevertheless, the stunning statistical significance of the CDF result throws into sharp relief the accelerating shift in the theory landscape from broad organizing principles like naturalness and unification to confronting empirical results at odds with the standard paradigm. As always, but particularly in this environment, new data will lead the way. With regard to the present discussion, new independent measurements of the $W$ mass are clearly warranted.

%% file: main.bbl
%apsrev4-2.bst 2019-01-14 (MD) hand-edited version of apsrev4-1.bst
%Control: key (0)
%Control: author (72) initials jnrlst
%Control: editor formatted (1) identically to author
%Control: production of article title (-1) disabled
%Control: page (0) single
%Control: year (1) truncated
%Control: production of eprint (0) enabled
\begin{thebibliography}{94}%
\makeatletter
\providecommand \@ifxundefined [1]{%
 \@ifx{#1\undefined}
}%
\providecommand \@ifnum [1]{%
 \ifnum #1\expandafter \@firstoftwo
 \else \expandafter \@secondoftwo
 \fi
}%
\providecommand \@ifx [1]{%
 \ifx #1\expandafter \@firstoftwo
 \else \expandafter \@secondoftwo
 \fi
}%
\providecommand \natexlab [1]{#1}%
\providecommand \enquote  [1]{``#1''}%
\providecommand \bibnamefont  [1]{#1}%
\providecommand \bibfnamefont [1]{#1}%
\providecommand \citenamefont [1]{#1}%
\providecommand \href@noop [0]{\@secondoftwo}%
\providecommand \href [0]{\begingroup \@sanitize@url \@href}%
\providecommand \@href[1]{\@@startlink{#1}\@@href}%
\providecommand \@@href[1]{\endgroup#1\@@endlink}%
\providecommand \@sanitize@url [0]{\catcode `\\12\catcode `\$12\catcode
  `\&12\catcode `\#12\catcode `\^12\catcode `\_12\catcode `\%12\relax}%
\providecommand \@@startlink[1]{}%
\providecommand \@@endlink[0]{}%
\providecommand \url  [0]{\begingroup\@sanitize@url \@url }%
\providecommand \@url [1]{\endgroup\@href {#1}{\urlprefix }}%
\providecommand \urlprefix  [0]{URL }%
\providecommand \Eprint [0]{\href }%
\providecommand \doibase [0]{https://doi.org/}%
\providecommand \selectlanguage [0]{\@gobble}%
\providecommand \bibinfo  [0]{\@secondoftwo}%
\providecommand \bibfield  [0]{\@secondoftwo}%
\providecommand \translation [1]{[#1]}%
\providecommand \BibitemOpen [0]{}%
\providecommand \bibitemStop [0]{}%
\providecommand \bibitemNoStop [0]{.\EOS\space}%
\providecommand \EOS [0]{\spacefactor3000\relax}%
\providecommand \BibitemShut  [1]{\csname bibitem#1\endcsname}%
\let\auto@bib@innerbib\@empty
%</preamble>
\bibitem [{\citenamefont {Aaltonen}\ \emph {et~al.}(2022)\citenamefont
  {Aaltonen} \emph {et~al.}}]{CDFWboson}%
  \BibitemOpen
  \bibfield  {author} {\bibinfo {author} {\bibfnamefont {T.}~\bibnamefont
  {Aaltonen}} \emph {et~al.} (\bibinfo {collaboration} {CDF}),\ }\href
  {https://doi.org/10.1126/science.abk1781} {\bibfield  {journal} {\bibinfo
  {journal} {Science}\ }\textbf {\bibinfo {volume} {376}},\ \bibinfo {pages}
  {170} (\bibinfo {year} {2022})}\BibitemShut {NoStop}%
\bibitem [{\citenamefont {Haller}\ \emph {et~al.}(2018)\citenamefont {Haller},
  \citenamefont {Hoecker}, \citenamefont {Kogler}, \citenamefont {M\"onig},
  \citenamefont {Peiffer},\ and\ \citenamefont {Stelzer}}]{Haller:2018nnx}%
  \BibitemOpen
  \bibfield  {author} {\bibinfo {author} {\bibfnamefont {J.}~\bibnamefont
  {Haller}}, \bibinfo {author} {\bibfnamefont {A.}~\bibnamefont {Hoecker}},
  \bibinfo {author} {\bibfnamefont {R.}~\bibnamefont {Kogler}}, \bibinfo
  {author} {\bibfnamefont {K.}~\bibnamefont {M\"onig}}, \bibinfo {author}
  {\bibfnamefont {T.}~\bibnamefont {Peiffer}},\ and\ \bibinfo {author}
  {\bibfnamefont {J.}~\bibnamefont {Stelzer}},\ }\href
  {https://doi.org/10.1140/epjc/s10052-018-6131-3} {\bibfield  {journal}
  {\bibinfo  {journal} {Eur. Phys. J. C}\ }\textbf {\bibinfo {volume} {78}},\
  \bibinfo {pages} {675} (\bibinfo {year} {2018})},\ \Eprint
  {https://arxiv.org/abs/1803.01853} {arXiv:1803.01853 [hep-ph]} \BibitemShut
  {NoStop}%
\bibitem [{\citenamefont {Zyla}\ \emph {et~al.}(2020)\citenamefont {Zyla} \emph
  {et~al.}}]{pdg2020}%
  \BibitemOpen
  \bibfield  {author} {\bibinfo {author} {\bibfnamefont {P.~A.}\ \bibnamefont
  {Zyla}} \emph {et~al.} (\bibinfo {collaboration} {Particle Data Group}),\
  }\href {https://doi.org/10.1093/ptep/ptaa104} {\bibfield  {journal} {\bibinfo
   {journal} {Prog. Theor. Exp. Phys.}\ }\textbf {\bibinfo {volume} {2020}},\
  \bibinfo {pages} {083C01} (\bibinfo {year} {2020})}\BibitemShut {NoStop}%
\bibitem [{\citenamefont {Aaltonen}\ \emph {et~al.}(2013)\citenamefont
  {Aaltonen} \emph {et~al.}}]{CDF:2013dpa}%
  \BibitemOpen
  \bibfield  {author} {\bibinfo {author} {\bibfnamefont {T.~A.}\ \bibnamefont
  {Aaltonen}} \emph {et~al.} (\bibinfo {collaboration} {CDF, D0}),\ }\href
  {https://doi.org/10.1103/PhysRevD.88.052018} {\bibfield  {journal} {\bibinfo
  {journal} {Phys. Rev. D}\ }\textbf {\bibinfo {volume} {88}},\ \bibinfo
  {pages} {052018} (\bibinfo {year} {2013})},\ \Eprint
  {https://arxiv.org/abs/1307.7627} {arXiv:1307.7627 [hep-ex]} \BibitemShut
  {NoStop}%
\bibitem [{\citenamefont {Aaboud}\ \emph
  {et~al.}(2018{\natexlab{a}})\citenamefont {Aaboud} \emph
  {et~al.}}]{ATLAS:2017rzl}%
  \BibitemOpen
  \bibfield  {author} {\bibinfo {author} {\bibfnamefont {M.}~\bibnamefont
  {Aaboud}} \emph {et~al.} (\bibinfo {collaboration} {ATLAS}),\ }\href
  {https://doi.org/10.1140/epjc/s10052-017-5475-4} {\bibfield  {journal}
  {\bibinfo  {journal} {Eur. Phys. J. C}\ }\textbf {\bibinfo {volume} {78}},\
  \bibinfo {pages} {110} (\bibinfo {year} {2018}{\natexlab{a}})},\ \bibinfo
  {note} {[Erratum: Eur.Phys.J.C 78, 898 (2018)]},\ \Eprint
  {https://arxiv.org/abs/1701.07240} {arXiv:1701.07240 [hep-ex]} \BibitemShut
  {NoStop}%
\bibitem [{\citenamefont {Aaij}\ \emph {et~al.}(2022)\citenamefont {Aaij} \emph
  {et~al.}}]{LHCb:2021bjt}%
  \BibitemOpen
  \bibfield  {author} {\bibinfo {author} {\bibfnamefont {R.}~\bibnamefont
  {Aaij}} \emph {et~al.} (\bibinfo {collaboration} {LHCb}),\ }\href
  {https://doi.org/10.1007/JHEP01(2022)036} {\bibfield  {journal} {\bibinfo
  {journal} {J. High Energy Phys.}\ }\textbf {\bibinfo {volume} {01}},\
  \bibinfo {pages} {036}},\ \Eprint {https://arxiv.org/abs/2109.01113}
  {arXiv:2109.01113 [hep-ex]} \BibitemShut {NoStop}%
\bibitem [{\citenamefont {Kanemura}\ and\ \citenamefont
  {Yagyu}(2022)}]{Kanemura:2022ahw}%
  \BibitemOpen
  \bibfield  {author} {\bibinfo {author} {\bibfnamefont {S.}~\bibnamefont
  {Kanemura}}\ and\ \bibinfo {author} {\bibfnamefont {K.}~\bibnamefont
  {Yagyu}},\ }\Eprint {https://arxiv.org/abs/2204.07511} {arXiv:2204.07511
  [hep-ph]}  (\bibinfo {year} {2022})\BibitemShut {NoStop}%
\bibitem [{\citenamefont {Nagao}\ \emph {et~al.}(2022)\citenamefont {Nagao},
  \citenamefont {Nomura},\ and\ \citenamefont {Okada}}]{Nagao:2022oin}%
  \BibitemOpen
  \bibfield  {author} {\bibinfo {author} {\bibfnamefont {K.~I.}\ \bibnamefont
  {Nagao}}, \bibinfo {author} {\bibfnamefont {T.}~\bibnamefont {Nomura}},\ and\
  \bibinfo {author} {\bibfnamefont {H.}~\bibnamefont {Okada}},\ }\Eprint
  {https://arxiv.org/abs/2204.07411} {arXiv:2204.07411 [hep-ph]}  (\bibinfo
  {year} {2022})\BibitemShut {NoStop}%
\bibitem [{\citenamefont {Perez}\ \emph {et~al.}(2022)\citenamefont {Perez},
  \citenamefont {Patel},\ and\ \citenamefont {Plascencia}}]{Perez:2022uil}%
  \BibitemOpen
  \bibfield  {author} {\bibinfo {author} {\bibfnamefont {P.~F.}\ \bibnamefont
  {Perez}}, \bibinfo {author} {\bibfnamefont {H.~H.}\ \bibnamefont {Patel}},\
  and\ \bibinfo {author} {\bibfnamefont {A.~D.}\ \bibnamefont {Plascencia}},\
  }\Eprint {https://arxiv.org/abs/2204.07144} {arXiv:2204.07144 [hep-ph]}
  (\bibinfo {year} {2022})\BibitemShut {NoStop}%
\bibitem [{\citenamefont {Ghoshal}\ \emph {et~al.}(2022)\citenamefont
  {Ghoshal}, \citenamefont {Okada}, \citenamefont {Okada}, \citenamefont
  {Raut}, \citenamefont {Shafi},\ and\ \citenamefont
  {Thapa}}]{Ghoshal:2022vzo}%
  \BibitemOpen
  \bibfield  {author} {\bibinfo {author} {\bibfnamefont {A.}~\bibnamefont
  {Ghoshal}}, \bibinfo {author} {\bibfnamefont {N.}~\bibnamefont {Okada}},
  \bibinfo {author} {\bibfnamefont {S.}~\bibnamefont {Okada}}, \bibinfo
  {author} {\bibfnamefont {D.}~\bibnamefont {Raut}}, \bibinfo {author}
  {\bibfnamefont {Q.}~\bibnamefont {Shafi}},\ and\ \bibinfo {author}
  {\bibfnamefont {A.}~\bibnamefont {Thapa}},\ }\Eprint
  {https://arxiv.org/abs/2204.07138} {arXiv:2204.07138 [hep-ph]}  (\bibinfo
  {year} {2022})\BibitemShut {NoStop}%
\bibitem [{\citenamefont {Kawamura}\ \emph {et~al.}(2022)\citenamefont
  {Kawamura}, \citenamefont {Okawa},\ and\ \citenamefont
  {Omura}}]{Kawamura:2022uft}%
  \BibitemOpen
  \bibfield  {author} {\bibinfo {author} {\bibfnamefont {J.}~\bibnamefont
  {Kawamura}}, \bibinfo {author} {\bibfnamefont {S.}~\bibnamefont {Okawa}},\
  and\ \bibinfo {author} {\bibfnamefont {Y.}~\bibnamefont {Omura}},\ }\Eprint
  {https://arxiv.org/abs/2204.07022} {arXiv:2204.07022 [hep-ph]}  (\bibinfo
  {year} {2022})\BibitemShut {NoStop}%
\bibitem [{\citenamefont {Zheng}\ \emph {et~al.}(2022)\citenamefont {Zheng},
  \citenamefont {Chen},\ and\ \citenamefont {Zhang}}]{Zheng:2022irz}%
  \BibitemOpen
  \bibfield  {author} {\bibinfo {author} {\bibfnamefont {M.-D.}\ \bibnamefont
  {Zheng}}, \bibinfo {author} {\bibfnamefont {F.-Z.}\ \bibnamefont {Chen}},\
  and\ \bibinfo {author} {\bibfnamefont {H.-H.}\ \bibnamefont {Zhang}},\
  }\Eprint {https://arxiv.org/abs/2204.06541} {arXiv:2204.06541 [hep-ph]}
  (\bibinfo {year} {2022})\BibitemShut {NoStop}%
\bibitem [{\citenamefont {Han}\ \emph {et~al.}(2022)\citenamefont {Han},
  \citenamefont {Wang}, \citenamefont {Wang}, \citenamefont {Yang},\ and\
  \citenamefont {Zhang}}]{Han:2022juu}%
  \BibitemOpen
  \bibfield  {author} {\bibinfo {author} {\bibfnamefont {X.-F.}\ \bibnamefont
  {Han}}, \bibinfo {author} {\bibfnamefont {F.}~\bibnamefont {Wang}}, \bibinfo
  {author} {\bibfnamefont {L.}~\bibnamefont {Wang}}, \bibinfo {author}
  {\bibfnamefont {J.~M.}\ \bibnamefont {Yang}},\ and\ \bibinfo {author}
  {\bibfnamefont {Y.}~\bibnamefont {Zhang}},\ }\Eprint
  {https://arxiv.org/abs/2204.06505} {arXiv:2204.06505 [hep-ph]}  (\bibinfo
  {year} {2022})\BibitemShut {NoStop}%
\bibitem [{\citenamefont {Ahn}\ \emph {et~al.}(2022)\citenamefont {Ahn},
  \citenamefont {Kang},\ and\ \citenamefont {Ramos}}]{Ahn:2022xeq}%
  \BibitemOpen
  \bibfield  {author} {\bibinfo {author} {\bibfnamefont {Y.~H.}\ \bibnamefont
  {Ahn}}, \bibinfo {author} {\bibfnamefont {S.~K.}\ \bibnamefont {Kang}},\ and\
  \bibinfo {author} {\bibfnamefont {R.}~\bibnamefont {Ramos}},\ }\Eprint
  {https://arxiv.org/abs/2204.06485} {arXiv:2204.06485 [hep-ph]}  (\bibinfo
  {year} {2022})\BibitemShut {NoStop}%
\bibitem [{\citenamefont {Balkin}\ \emph {et~al.}(2022)\citenamefont {Balkin},
  \citenamefont {Madge}, \citenamefont {Menzo}, \citenamefont {Perez},
  \citenamefont {Soreq},\ and\ \citenamefont {Zupan}}]{Balkin:2022glu}%
  \BibitemOpen
  \bibfield  {author} {\bibinfo {author} {\bibfnamefont {R.}~\bibnamefont
  {Balkin}}, \bibinfo {author} {\bibfnamefont {E.}~\bibnamefont {Madge}},
  \bibinfo {author} {\bibfnamefont {T.}~\bibnamefont {Menzo}}, \bibinfo
  {author} {\bibfnamefont {G.}~\bibnamefont {Perez}}, \bibinfo {author}
  {\bibfnamefont {Y.}~\bibnamefont {Soreq}},\ and\ \bibinfo {author}
  {\bibfnamefont {J.}~\bibnamefont {Zupan}},\ }\Eprint
  {https://arxiv.org/abs/2204.05992} {arXiv:2204.05992 [hep-ph]}  (\bibinfo
  {year} {2022})\BibitemShut {NoStop}%
\bibitem [{\citenamefont {Biek\"otter}\ \emph {et~al.}(2022)\citenamefont
  {Biek\"otter}, \citenamefont {Heinemeyer},\ and\ \citenamefont
  {Weiglein}}]{Biekotter:2022abc}%
  \BibitemOpen
  \bibfield  {author} {\bibinfo {author} {\bibfnamefont {T.}~\bibnamefont
  {Biek\"otter}}, \bibinfo {author} {\bibfnamefont {S.}~\bibnamefont
  {Heinemeyer}},\ and\ \bibinfo {author} {\bibfnamefont {G.}~\bibnamefont
  {Weiglein}},\ }\Eprint {https://arxiv.org/abs/2204.05975} {arXiv:2204.05975
  [hep-ph]}  (\bibinfo {year} {2022})\BibitemShut {NoStop}%
\bibitem [{\citenamefont {Endo}\ and\ \citenamefont
  {Mishima}(2022)}]{Endo:2022kiw}%
  \BibitemOpen
  \bibfield  {author} {\bibinfo {author} {\bibfnamefont {M.}~\bibnamefont
  {Endo}}\ and\ \bibinfo {author} {\bibfnamefont {S.}~\bibnamefont {Mishima}},\
  }\Eprint {https://arxiv.org/abs/2204.05965} {arXiv:2204.05965 [hep-ph]}
  (\bibinfo {year} {2022})\BibitemShut {NoStop}%
\bibitem [{\citenamefont {Crivellin}\ \emph {et~al.}(2022)\citenamefont
  {Crivellin}, \citenamefont {Kirk}, \citenamefont {Kitahara},\ and\
  \citenamefont {Mescia}}]{Crivellin:2022fdf}%
  \BibitemOpen
  \bibfield  {author} {\bibinfo {author} {\bibfnamefont {A.}~\bibnamefont
  {Crivellin}}, \bibinfo {author} {\bibfnamefont {M.}~\bibnamefont {Kirk}},
  \bibinfo {author} {\bibfnamefont {T.}~\bibnamefont {Kitahara}},\ and\
  \bibinfo {author} {\bibfnamefont {F.}~\bibnamefont {Mescia}},\ }\Eprint
  {https://arxiv.org/abs/2204.05962} {arXiv:2204.05962 [hep-ph]}  (\bibinfo
  {year} {2022})\BibitemShut {NoStop}%
\bibitem [{\citenamefont {Cheung}\ \emph {et~al.}(2022)\citenamefont {Cheung},
  \citenamefont {Keung},\ and\ \citenamefont {Tseng}}]{Cheung:2022zsb}%
  \BibitemOpen
  \bibfield  {author} {\bibinfo {author} {\bibfnamefont {K.}~\bibnamefont
  {Cheung}}, \bibinfo {author} {\bibfnamefont {W.-Y.}\ \bibnamefont {Keung}},\
  and\ \bibinfo {author} {\bibfnamefont {P.-Y.}\ \bibnamefont {Tseng}},\
  }\Eprint {https://arxiv.org/abs/2204.05942} {arXiv:2204.05942 [hep-ph]}
  (\bibinfo {year} {2022})\BibitemShut {NoStop}%
\bibitem [{\citenamefont {Du}\ \emph {et~al.}(2022)\citenamefont {Du},
  \citenamefont {Li}, \citenamefont {Wang},\ and\ \citenamefont
  {Zhang}}]{Du:2022brr}%
  \BibitemOpen
  \bibfield  {author} {\bibinfo {author} {\bibfnamefont {X.~K.}\ \bibnamefont
  {Du}}, \bibinfo {author} {\bibfnamefont {Z.}~\bibnamefont {Li}}, \bibinfo
  {author} {\bibfnamefont {F.}~\bibnamefont {Wang}},\ and\ \bibinfo {author}
  {\bibfnamefont {Y.~K.}\ \bibnamefont {Zhang}},\ }\Eprint
  {https://arxiv.org/abs/2204.05760} {arXiv:2204.05760 [hep-ph]}  (\bibinfo
  {year} {2022})\BibitemShut {NoStop}%
\bibitem [{\citenamefont {Heo}\ \emph {et~al.}(2022)\citenamefont {Heo},
  \citenamefont {Jung},\ and\ \citenamefont {Lee}}]{Heo:2022dey}%
  \BibitemOpen
  \bibfield  {author} {\bibinfo {author} {\bibfnamefont {Y.}~\bibnamefont
  {Heo}}, \bibinfo {author} {\bibfnamefont {D.-W.}\ \bibnamefont {Jung}},\ and\
  \bibinfo {author} {\bibfnamefont {J.~S.}\ \bibnamefont {Lee}},\ }\Eprint
  {https://arxiv.org/abs/2204.05728} {arXiv:2204.05728 [hep-ph]}  (\bibinfo
  {year} {2022})\BibitemShut {NoStop}%
\bibitem [{\citenamefont {Babu}\ \emph {et~al.}(2022)\citenamefont {Babu},
  \citenamefont {Jana},\ and\ \citenamefont {K.}}]{Babu:2022pdn}%
  \BibitemOpen
  \bibfield  {author} {\bibinfo {author} {\bibfnamefont {K.~S.}\ \bibnamefont
  {Babu}}, \bibinfo {author} {\bibfnamefont {S.}~\bibnamefont {Jana}},\ and\
  \bibinfo {author} {\bibfnamefont {V.~P.}\ \bibnamefont {K.}},\ }\Eprint
  {https://arxiv.org/abs/2204.05303} {arXiv:2204.05303 [hep-ph]}  (\bibinfo
  {year} {2022})\BibitemShut {NoStop}%
\bibitem [{\citenamefont {Heckman}(2022)}]{Heckman:2022the}%
  \BibitemOpen
  \bibfield  {author} {\bibinfo {author} {\bibfnamefont {J.~J.}\ \bibnamefont
  {Heckman}},\ }\Eprint {https://arxiv.org/abs/2204.05302} {arXiv:2204.05302
  [hep-ph]}  (\bibinfo {year} {2022})\BibitemShut {NoStop}%
\bibitem [{\citenamefont {Gu}\ \emph {et~al.}(2022)\citenamefont {Gu},
  \citenamefont {Liu}, \citenamefont {Ma},\ and\ \citenamefont
  {Shu}}]{Gu:2022htv}%
  \BibitemOpen
  \bibfield  {author} {\bibinfo {author} {\bibfnamefont {J.}~\bibnamefont
  {Gu}}, \bibinfo {author} {\bibfnamefont {Z.}~\bibnamefont {Liu}}, \bibinfo
  {author} {\bibfnamefont {T.}~\bibnamefont {Ma}},\ and\ \bibinfo {author}
  {\bibfnamefont {J.}~\bibnamefont {Shu}},\ }\Eprint
  {https://arxiv.org/abs/2204.05296} {arXiv:2204.05296 [hep-ph]}  (\bibinfo
  {year} {2022})\BibitemShut {NoStop}%
\bibitem [{\citenamefont {Athron}\ \emph
  {et~al.}(2022{\natexlab{a}})\citenamefont {Athron}, \citenamefont {Bach},
  \citenamefont {Jacob}, \citenamefont {Kotlarski}, \citenamefont
  {St\"ockinger},\ and\ \citenamefont {Voigt}}]{Athron:2022isz}%
  \BibitemOpen
  \bibfield  {author} {\bibinfo {author} {\bibfnamefont {P.}~\bibnamefont
  {Athron}}, \bibinfo {author} {\bibfnamefont {M.}~\bibnamefont {Bach}},
  \bibinfo {author} {\bibfnamefont {D.~H.~J.}\ \bibnamefont {Jacob}}, \bibinfo
  {author} {\bibfnamefont {W.}~\bibnamefont {Kotlarski}}, \bibinfo {author}
  {\bibfnamefont {D.}~\bibnamefont {St\"ockinger}},\ and\ \bibinfo {author}
  {\bibfnamefont {A.}~\bibnamefont {Voigt}},\ }\Eprint
  {https://arxiv.org/abs/2204.05285} {arXiv:2204.05285 [hep-ph]}  (\bibinfo
  {year} {2022}{\natexlab{a}})\BibitemShut {NoStop}%
\bibitem [{\citenamefont {Di~Luzio}\ \emph {et~al.}(2022)\citenamefont
  {Di~Luzio}, \citenamefont {Gr\"ober},\ and\ \citenamefont
  {Paradisi}}]{DiLuzio:2022xns}%
  \BibitemOpen
  \bibfield  {author} {\bibinfo {author} {\bibfnamefont {L.}~\bibnamefont
  {Di~Luzio}}, \bibinfo {author} {\bibfnamefont {R.}~\bibnamefont {Gr\"ober}},\
  and\ \bibinfo {author} {\bibfnamefont {P.}~\bibnamefont {Paradisi}},\
  }\Eprint {https://arxiv.org/abs/2204.05284} {arXiv:2204.05284 [hep-ph]}
  (\bibinfo {year} {2022})\BibitemShut {NoStop}%
\bibitem [{\citenamefont {Asadi}\ \emph {et~al.}(2022)\citenamefont {Asadi},
  \citenamefont {Cesarotti}, \citenamefont {Fraser}, \citenamefont {Homiller},\
  and\ \citenamefont {Parikh}}]{Asadi:2022xiy}%
  \BibitemOpen
  \bibfield  {author} {\bibinfo {author} {\bibfnamefont {P.}~\bibnamefont
  {Asadi}}, \bibinfo {author} {\bibfnamefont {C.}~\bibnamefont {Cesarotti}},
  \bibinfo {author} {\bibfnamefont {K.}~\bibnamefont {Fraser}}, \bibinfo
  {author} {\bibfnamefont {S.}~\bibnamefont {Homiller}},\ and\ \bibinfo
  {author} {\bibfnamefont {A.}~\bibnamefont {Parikh}},\ }\Eprint
  {https://arxiv.org/abs/2204.05283} {arXiv:2204.05283 [hep-ph]}  (\bibinfo
  {year} {2022})\BibitemShut {NoStop}%
\bibitem [{\citenamefont {Bahl}\ \emph {et~al.}(2022)\citenamefont {Bahl},
  \citenamefont {Braathen},\ and\ \citenamefont {Weiglein}}]{Bahl:2022xzi}%
  \BibitemOpen
  \bibfield  {author} {\bibinfo {author} {\bibfnamefont {H.}~\bibnamefont
  {Bahl}}, \bibinfo {author} {\bibfnamefont {J.}~\bibnamefont {Braathen}},\
  and\ \bibinfo {author} {\bibfnamefont {G.}~\bibnamefont {Weiglein}},\
  }\Eprint {https://arxiv.org/abs/2204.05269} {arXiv:2204.05269 [hep-ph]}
  (\bibinfo {year} {2022})\BibitemShut {NoStop}%
\bibitem [{\citenamefont {Paul}\ and\ \citenamefont
  {Valli}(2022)}]{Paul:2022dds}%
  \BibitemOpen
  \bibfield  {author} {\bibinfo {author} {\bibfnamefont {A.}~\bibnamefont
  {Paul}}\ and\ \bibinfo {author} {\bibfnamefont {M.}~\bibnamefont {Valli}},\
  }\Eprint {https://arxiv.org/abs/2204.05267} {arXiv:2204.05267 [hep-ph]}
  (\bibinfo {year} {2022})\BibitemShut {NoStop}%
\bibitem [{\citenamefont {Bagnaschi}\ \emph {et~al.}(2022)\citenamefont
  {Bagnaschi}, \citenamefont {Ellis}, \citenamefont {Madigan}, \citenamefont
  {Mimasu}, \citenamefont {Sanz},\ and\ \citenamefont
  {You}}]{Bagnaschi:2022whn}%
  \BibitemOpen
  \bibfield  {author} {\bibinfo {author} {\bibfnamefont {E.}~\bibnamefont
  {Bagnaschi}}, \bibinfo {author} {\bibfnamefont {J.}~\bibnamefont {Ellis}},
  \bibinfo {author} {\bibfnamefont {M.}~\bibnamefont {Madigan}}, \bibinfo
  {author} {\bibfnamefont {K.}~\bibnamefont {Mimasu}}, \bibinfo {author}
  {\bibfnamefont {V.}~\bibnamefont {Sanz}},\ and\ \bibinfo {author}
  {\bibfnamefont {T.}~\bibnamefont {You}},\ }\Eprint
  {https://arxiv.org/abs/2204.05260} {arXiv:2204.05260 [hep-ph]}  (\bibinfo
  {year} {2022})\BibitemShut {NoStop}%
\bibitem [{\citenamefont {Song}\ \emph {et~al.}(2022)\citenamefont {Song},
  \citenamefont {Su},\ and\ \citenamefont {Zhang}}]{Song:2022xts}%
  \BibitemOpen
  \bibfield  {author} {\bibinfo {author} {\bibfnamefont {H.}~\bibnamefont
  {Song}}, \bibinfo {author} {\bibfnamefont {W.}~\bibnamefont {Su}},\ and\
  \bibinfo {author} {\bibfnamefont {M.}~\bibnamefont {Zhang}},\ }\Eprint
  {https://arxiv.org/abs/2204.05085} {arXiv:2204.05085 [hep-ph]}  (\bibinfo
  {year} {2022})\BibitemShut {NoStop}%
\bibitem [{\citenamefont {Cheng}\ \emph {et~al.}(2022)\citenamefont {Cheng},
  \citenamefont {He}, \citenamefont {Huang},\ and\ \citenamefont
  {Li}}]{Cheng:2022jyi}%
  \BibitemOpen
  \bibfield  {author} {\bibinfo {author} {\bibfnamefont {Y.}~\bibnamefont
  {Cheng}}, \bibinfo {author} {\bibfnamefont {X.-G.}\ \bibnamefont {He}},
  \bibinfo {author} {\bibfnamefont {Z.-L.}\ \bibnamefont {Huang}},\ and\
  \bibinfo {author} {\bibfnamefont {M.-W.}\ \bibnamefont {Li}},\ }\Eprint
  {https://arxiv.org/abs/2204.05031} {arXiv:2204.05031 [hep-ph]}  (\bibinfo
  {year} {2022})\BibitemShut {NoStop}%
\bibitem [{\citenamefont {Lee}\ and\ \citenamefont
  {Yamashita}(2022)}]{Lee:2022nqz}%
  \BibitemOpen
  \bibfield  {author} {\bibinfo {author} {\bibfnamefont {H.~M.}\ \bibnamefont
  {Lee}}\ and\ \bibinfo {author} {\bibfnamefont {K.}~\bibnamefont
  {Yamashita}},\ }\Eprint {https://arxiv.org/abs/2204.05024} {arXiv:2204.05024
  [hep-ph]}  (\bibinfo {year} {2022})\BibitemShut {NoStop}%
\bibitem [{\citenamefont {Liu}\ \emph {et~al.}(2022)\citenamefont {Liu},
  \citenamefont {Guo}, \citenamefont {Zhu},\ and\ \citenamefont
  {Li}}]{Liu:2022jdq}%
  \BibitemOpen
  \bibfield  {author} {\bibinfo {author} {\bibfnamefont {X.}~\bibnamefont
  {Liu}}, \bibinfo {author} {\bibfnamefont {S.-Y.}\ \bibnamefont {Guo}},
  \bibinfo {author} {\bibfnamefont {B.}~\bibnamefont {Zhu}},\ and\ \bibinfo
  {author} {\bibfnamefont {Y.}~\bibnamefont {Li}},\ }\Eprint
  {https://arxiv.org/abs/2204.04834} {arXiv:2204.04834 [hep-ph]}  (\bibinfo
  {year} {2022})\BibitemShut {NoStop}%
\bibitem [{\citenamefont {Fan}\ \emph {et~al.}(2022{\natexlab{a}})\citenamefont
  {Fan}, \citenamefont {Li}, \citenamefont {Liu},\ and\ \citenamefont
  {Lyu}}]{Fan:2022yly}%
  \BibitemOpen
  \bibfield  {author} {\bibinfo {author} {\bibfnamefont {J.}~\bibnamefont
  {Fan}}, \bibinfo {author} {\bibfnamefont {L.}~\bibnamefont {Li}}, \bibinfo
  {author} {\bibfnamefont {T.}~\bibnamefont {Liu}},\ and\ \bibinfo {author}
  {\bibfnamefont {K.-F.}\ \bibnamefont {Lyu}},\ }\Eprint
  {https://arxiv.org/abs/2204.04805} {arXiv:2204.04805 [hep-ph]}  (\bibinfo
  {year} {2022}{\natexlab{a}})\BibitemShut {NoStop}%
\bibitem [{\citenamefont {Sakurai}\ \emph {et~al.}(2022)\citenamefont
  {Sakurai}, \citenamefont {Takahashi},\ and\ \citenamefont
  {Yin}}]{Sakurai:2022hwh}%
  \BibitemOpen
  \bibfield  {author} {\bibinfo {author} {\bibfnamefont {K.}~\bibnamefont
  {Sakurai}}, \bibinfo {author} {\bibfnamefont {F.}~\bibnamefont {Takahashi}},\
  and\ \bibinfo {author} {\bibfnamefont {W.}~\bibnamefont {Yin}},\ }\Eprint
  {https://arxiv.org/abs/2204.04770} {arXiv:2204.04770 [hep-ph]}  (\bibinfo
  {year} {2022})\BibitemShut {NoStop}%
\bibitem [{\citenamefont {Fan}\ \emph {et~al.}(2022{\natexlab{b}})\citenamefont
  {Fan}, \citenamefont {Tang}, \citenamefont {Tsai},\ and\ \citenamefont
  {Wu}}]{Fan:2022dck}%
  \BibitemOpen
  \bibfield  {author} {\bibinfo {author} {\bibfnamefont {Y.-Z.}\ \bibnamefont
  {Fan}}, \bibinfo {author} {\bibfnamefont {T.-P.}\ \bibnamefont {Tang}},
  \bibinfo {author} {\bibfnamefont {Y.-L.~S.}\ \bibnamefont {Tsai}},\ and\
  \bibinfo {author} {\bibfnamefont {L.}~\bibnamefont {Wu}},\ }\Eprint
  {https://arxiv.org/abs/2204.03693} {arXiv:2204.03693 [hep-ph]}  (\bibinfo
  {year} {2022}{\natexlab{b}})\BibitemShut {NoStop}%
\bibitem [{\citenamefont {Lu}\ \emph {et~al.}(2022)\citenamefont {Lu},
  \citenamefont {Wu}, \citenamefont {Wu},\ and\ \citenamefont
  {Zhu}}]{Lu:2022bgw}%
  \BibitemOpen
  \bibfield  {author} {\bibinfo {author} {\bibfnamefont {C.-T.}\ \bibnamefont
  {Lu}}, \bibinfo {author} {\bibfnamefont {L.}~\bibnamefont {Wu}}, \bibinfo
  {author} {\bibfnamefont {Y.}~\bibnamefont {Wu}},\ and\ \bibinfo {author}
  {\bibfnamefont {B.}~\bibnamefont {Zhu}},\ }\Eprint
  {https://arxiv.org/abs/2204.03796} {arXiv:2204.03796 [hep-ph]}  (\bibinfo
  {year} {2022})\BibitemShut {NoStop}%
\bibitem [{\citenamefont {Athron}\ \emph
  {et~al.}(2022{\natexlab{b}})\citenamefont {Athron}, \citenamefont {Fowlie},
  \citenamefont {Lu}, \citenamefont {Wu}, \citenamefont {Wu},\ and\
  \citenamefont {Zhu}}]{Athron:2022qpo}%
  \BibitemOpen
  \bibfield  {author} {\bibinfo {author} {\bibfnamefont {P.}~\bibnamefont
  {Athron}}, \bibinfo {author} {\bibfnamefont {A.}~\bibnamefont {Fowlie}},
  \bibinfo {author} {\bibfnamefont {C.-T.}\ \bibnamefont {Lu}}, \bibinfo
  {author} {\bibfnamefont {L.}~\bibnamefont {Wu}}, \bibinfo {author}
  {\bibfnamefont {Y.}~\bibnamefont {Wu}},\ and\ \bibinfo {author}
  {\bibfnamefont {B.}~\bibnamefont {Zhu}},\ }\Eprint
  {https://arxiv.org/abs/2204.03996} {arXiv:2204.03996 [hep-ph]}  (\bibinfo
  {year} {2022}{\natexlab{b}})\BibitemShut {NoStop}%
\bibitem [{\citenamefont {Yuan}\ \emph {et~al.}(2022)\citenamefont {Yuan},
  \citenamefont {Zu}, \citenamefont {Feng},\ and\ \citenamefont
  {Cai}}]{Yuan:2022cpw}%
  \BibitemOpen
  \bibfield  {author} {\bibinfo {author} {\bibfnamefont {G.-W.}\ \bibnamefont
  {Yuan}}, \bibinfo {author} {\bibfnamefont {L.}~\bibnamefont {Zu}}, \bibinfo
  {author} {\bibfnamefont {L.}~\bibnamefont {Feng}},\ and\ \bibinfo {author}
  {\bibfnamefont {Y.-F.}\ \bibnamefont {Cai}},\ }\Eprint
  {https://arxiv.org/abs/2204.04183} {arXiv:2204.04183 [hep-ph]}  (\bibinfo
  {year} {2022})\BibitemShut {NoStop}%
\bibitem [{\citenamefont {Strumia}(2022)}]{Strumia:2022qkt}%
  \BibitemOpen
  \bibfield  {author} {\bibinfo {author} {\bibfnamefont {A.}~\bibnamefont
  {Strumia}},\ }\Eprint {https://arxiv.org/abs/2204.04191} {arXiv:2204.04191
  [hep-ph]}  (\bibinfo {year} {2022})\BibitemShut {NoStop}%
\bibitem [{\citenamefont {Yang}\ and\ \citenamefont
  {Zhang}(2022)}]{Yang:2022gvz}%
  \BibitemOpen
  \bibfield  {author} {\bibinfo {author} {\bibfnamefont {J.~M.}\ \bibnamefont
  {Yang}}\ and\ \bibinfo {author} {\bibfnamefont {Y.}~\bibnamefont {Zhang}},\
  }\Eprint {https://arxiv.org/abs/2204.04202} {arXiv:2204.04202 [hep-ph]}
  (\bibinfo {year} {2022})\BibitemShut {NoStop}%
\bibitem [{\citenamefont {de~Blas}\ \emph {et~al.}(2022)\citenamefont
  {de~Blas}, \citenamefont {Pierini}, \citenamefont {Reina},\ and\
  \citenamefont {Silvestrini}}]{deBlas:2022hdk}%
  \BibitemOpen
  \bibfield  {author} {\bibinfo {author} {\bibfnamefont {J.}~\bibnamefont
  {de~Blas}}, \bibinfo {author} {\bibfnamefont {M.}~\bibnamefont {Pierini}},
  \bibinfo {author} {\bibfnamefont {L.}~\bibnamefont {Reina}},\ and\ \bibinfo
  {author} {\bibfnamefont {L.}~\bibnamefont {Silvestrini}},\ }\Eprint
  {https://arxiv.org/abs/2204.04204} {arXiv:2204.04204 [hep-ph]}  (\bibinfo
  {year} {2022})\BibitemShut {NoStop}%
\bibitem [{\citenamefont {Peskin}\ and\ \citenamefont
  {Takeuchi}(1990)}]{Peskin:1990zt}%
  \BibitemOpen
  \bibfield  {author} {\bibinfo {author} {\bibfnamefont {M.~E.}\ \bibnamefont
  {Peskin}}\ and\ \bibinfo {author} {\bibfnamefont {T.}~\bibnamefont
  {Takeuchi}},\ }\href {https://doi.org/10.1103/PhysRevLett.65.964} {\bibfield
  {journal} {\bibinfo  {journal} {Phys. Rev. Lett.}\ }\textbf {\bibinfo
  {volume} {65}},\ \bibinfo {pages} {964} (\bibinfo {year} {1990})}\BibitemShut
  {NoStop}%
\bibitem [{\citenamefont {Peskin}\ and\ \citenamefont
  {Takeuchi}(1992)}]{Peskin:1991sw}%
  \BibitemOpen
  \bibfield  {author} {\bibinfo {author} {\bibfnamefont {M.~E.}\ \bibnamefont
  {Peskin}}\ and\ \bibinfo {author} {\bibfnamefont {T.}~\bibnamefont
  {Takeuchi}},\ }\href {https://doi.org/10.1103/PhysRevD.46.381} {\bibfield
  {journal} {\bibinfo  {journal} {Phys. Rev. D}\ }\textbf {\bibinfo {volume}
  {46}},\ \bibinfo {pages} {381} (\bibinfo {year} {1992})}\BibitemShut
  {NoStop}%
\bibitem [{\citenamefont {Crivellin}\ \emph {et~al.}(2020)\citenamefont
  {Crivellin}, \citenamefont {M\"uller},\ and\ \citenamefont
  {Saturnino}}]{Crivellin:2020ukd}%
  \BibitemOpen
  \bibfield  {author} {\bibinfo {author} {\bibfnamefont {A.}~\bibnamefont
  {Crivellin}}, \bibinfo {author} {\bibfnamefont {D.}~\bibnamefont
  {M\"uller}},\ and\ \bibinfo {author} {\bibfnamefont {F.}~\bibnamefont
  {Saturnino}},\ }\href {https://doi.org/10.1007/JHEP11(2020)094} {\bibfield
  {journal} {\bibinfo  {journal} {J. High Energy Phys.}\ }\textbf {\bibinfo
  {volume} {11}},\ \bibinfo {pages} {094}},\ \Eprint
  {https://arxiv.org/abs/2006.10758} {arXiv:2006.10758 [hep-ph]} \BibitemShut
  {NoStop}%
\bibitem [{\citenamefont {Gerbush}\ \emph {et~al.}(2008)\citenamefont
  {Gerbush}, \citenamefont {Khoo}, \citenamefont {Phalen}, \citenamefont
  {Pierce},\ and\ \citenamefont {Tucker-Smith}}]{Gerbush:2007fe}%
  \BibitemOpen
  \bibfield  {author} {\bibinfo {author} {\bibfnamefont {M.}~\bibnamefont
  {Gerbush}}, \bibinfo {author} {\bibfnamefont {T.~J.}\ \bibnamefont {Khoo}},
  \bibinfo {author} {\bibfnamefont {D.~J.}\ \bibnamefont {Phalen}}, \bibinfo
  {author} {\bibfnamefont {A.}~\bibnamefont {Pierce}},\ and\ \bibinfo {author}
  {\bibfnamefont {D.}~\bibnamefont {Tucker-Smith}},\ }\href
  {https://doi.org/10.1103/PhysRevD.77.095003} {\bibfield  {journal} {\bibinfo
  {journal} {Phys. Rev. D}\ }\textbf {\bibinfo {volume} {77}},\ \bibinfo
  {pages} {095003} (\bibinfo {year} {2008})},\ \Eprint
  {https://arxiv.org/abs/0710.3133} {arXiv:0710.3133 [hep-ph]} \BibitemShut
  {NoStop}%
\bibitem [{\citenamefont {Choi}\ \emph {et~al.}(2009)\citenamefont {Choi},
  \citenamefont {Drees}, \citenamefont {Kalinowski}, \citenamefont {Kim},
  \citenamefont {Popenda},\ and\ \citenamefont {Zerwas}}]{Choi:2008ub}%
  \BibitemOpen
  \bibfield  {author} {\bibinfo {author} {\bibfnamefont {S.~Y.}\ \bibnamefont
  {Choi}}, \bibinfo {author} {\bibfnamefont {M.}~\bibnamefont {Drees}},
  \bibinfo {author} {\bibfnamefont {J.}~\bibnamefont {Kalinowski}}, \bibinfo
  {author} {\bibfnamefont {J.~M.}\ \bibnamefont {Kim}}, \bibinfo {author}
  {\bibfnamefont {E.}~\bibnamefont {Popenda}},\ and\ \bibinfo {author}
  {\bibfnamefont {P.~M.}\ \bibnamefont {Zerwas}},\ }\href
  {https://doi.org/10.1016/j.physletb.2009.01.040} {\bibfield  {journal}
  {\bibinfo  {journal} {Phys. Lett. B}\ }\textbf {\bibinfo {volume} {672}},\
  \bibinfo {pages} {246} (\bibinfo {year} {2009})},\ \Eprint
  {https://arxiv.org/abs/0812.3586} {arXiv:0812.3586 [hep-ph]} \BibitemShut
  {NoStop}%
\bibitem [{\citenamefont {Plehn}\ and\ \citenamefont
  {Tait}(2009)}]{Plehn:2008ae}%
  \BibitemOpen
  \bibfield  {author} {\bibinfo {author} {\bibfnamefont {T.}~\bibnamefont
  {Plehn}}\ and\ \bibinfo {author} {\bibfnamefont {T.~M.~P.}\ \bibnamefont
  {Tait}},\ }\href {https://doi.org/10.1088/0954-3899/36/7/075001} {\bibfield
  {journal} {\bibinfo  {journal} {J. Phys. G}\ }\textbf {\bibinfo {volume}
  {36}},\ \bibinfo {pages} {075001} (\bibinfo {year} {2009})},\ \Eprint
  {https://arxiv.org/abs/0810.3919} {arXiv:0810.3919 [hep-ph]} \BibitemShut
  {NoStop}%
\bibitem [{\citenamefont {Chen}\ \emph {et~al.}(2015)\citenamefont {Chen},
  \citenamefont {Freitas}, \citenamefont {Han},\ and\ \citenamefont
  {Lee}}]{Chen:2014haa}%
  \BibitemOpen
  \bibfield  {author} {\bibinfo {author} {\bibfnamefont {C.-Y.}\ \bibnamefont
  {Chen}}, \bibinfo {author} {\bibfnamefont {A.}~\bibnamefont {Freitas}},
  \bibinfo {author} {\bibfnamefont {T.}~\bibnamefont {Han}},\ and\ \bibinfo
  {author} {\bibfnamefont {K.~S.~M.}\ \bibnamefont {Lee}},\ }\href
  {https://doi.org/10.1007/JHEP05(2015)135} {\bibfield  {journal} {\bibinfo
  {journal} {J. High Energy Phys.}\ }\textbf {\bibinfo {volume} {05}},\
  \bibinfo {pages} {135}},\ \Eprint {https://arxiv.org/abs/1410.8113}
  {arXiv:1410.8113 [hep-ph]} \BibitemShut {NoStop}%
\bibitem [{\citenamefont {Carpenter}\ and\ \citenamefont
  {Colburn}(2015)}]{Carpenter:2015gua}%
  \BibitemOpen
  \bibfield  {author} {\bibinfo {author} {\bibfnamefont {L.~M.}\ \bibnamefont
  {Carpenter}}\ and\ \bibinfo {author} {\bibfnamefont {R.}~\bibnamefont
  {Colburn}},\ }\href {https://doi.org/10.1007/JHEP12(2015)151} {\bibfield
  {journal} {\bibinfo  {journal} {J. High Energy Phys.}\ }\textbf {\bibinfo
  {volume} {12}},\ \bibinfo {pages} {151}},\ \Eprint
  {https://arxiv.org/abs/1509.07869} {arXiv:1509.07869 [hep-ph]} \BibitemShut
  {NoStop}%
\bibitem [{\citenamefont {Carpenter}\ \emph {et~al.}(2020)\citenamefont
  {Carpenter}, \citenamefont {Murphy},\ and\ \citenamefont
  {Smylie}}]{Carpenter:2020hyz}%
  \BibitemOpen
  \bibfield  {author} {\bibinfo {author} {\bibfnamefont {L.~M.}\ \bibnamefont
  {Carpenter}}, \bibinfo {author} {\bibfnamefont {T.}~\bibnamefont {Murphy}},\
  and\ \bibinfo {author} {\bibfnamefont {M.~J.}\ \bibnamefont {Smylie}},\
  }\href {https://doi.org/10.1007/JHEP11(2020)024} {\bibfield  {journal}
  {\bibinfo  {journal} {J. High Energy Phys.}\ }\textbf {\bibinfo {volume}
  {11}},\ \bibinfo {pages} {024}},\ \Eprint {https://arxiv.org/abs/2006.15217}
  {arXiv:2006.15217 [hep-ph]} \BibitemShut {NoStop}%
\bibitem [{\citenamefont {Carpenter}\ and\ \citenamefont
  {Murphy}(2021)}]{Carpenter:2020evo}%
  \BibitemOpen
  \bibfield  {author} {\bibinfo {author} {\bibfnamefont {L.~M.}\ \bibnamefont
  {Carpenter}}\ and\ \bibinfo {author} {\bibfnamefont {T.}~\bibnamefont
  {Murphy}},\ }\href {https://doi.org/10.1007/JHEP05(2021)079} {\bibfield
  {journal} {\bibinfo  {journal} {J. High Energy Phys.}\ }\textbf {\bibinfo
  {volume} {05}},\ \bibinfo {pages} {079}},\ \Eprint
  {https://arxiv.org/abs/2012.15771} {arXiv:2012.15771 [hep-ph]} \BibitemShut
  {NoStop}%
\bibitem [{\citenamefont {Carpenter}\ \emph
  {et~al.}(2022{\natexlab{a}})\citenamefont {Carpenter}, \citenamefont
  {Murphy},\ and\ \citenamefont {Smylie}}]{Carpenter:2021vga}%
  \BibitemOpen
  \bibfield  {author} {\bibinfo {author} {\bibfnamefont {L.~M.}\ \bibnamefont
  {Carpenter}}, \bibinfo {author} {\bibfnamefont {T.}~\bibnamefont {Murphy}},\
  and\ \bibinfo {author} {\bibfnamefont {M.~J.}\ \bibnamefont {Smylie}},\
  }\href {https://doi.org/10.1007/JHEP01(2022)047} {\bibfield  {journal}
  {\bibinfo  {journal} {J. High Energy Phys.}\ }\textbf {\bibinfo {volume}
  {01}},\ \bibinfo {pages} {047}},\ \Eprint {https://arxiv.org/abs/2107.13565}
  {arXiv:2107.13565 [hep-ph]} \BibitemShut {NoStop}%
\bibitem [{\citenamefont {Chen}\ \emph {et~al.}(2009)\citenamefont {Chen},
  \citenamefont {Klemm}, \citenamefont {Rentala},\ and\ \citenamefont
  {Wang}}]{Chen:2008hh}%
  \BibitemOpen
  \bibfield  {author} {\bibinfo {author} {\bibfnamefont {C.-R.}\ \bibnamefont
  {Chen}}, \bibinfo {author} {\bibfnamefont {W.}~\bibnamefont {Klemm}},
  \bibinfo {author} {\bibfnamefont {V.}~\bibnamefont {Rentala}},\ and\ \bibinfo
  {author} {\bibfnamefont {K.}~\bibnamefont {Wang}},\ }\href
  {https://doi.org/10.1103/PhysRevD.79.054002} {\bibfield  {journal} {\bibinfo
  {journal} {Phys. Rev. D}\ }\textbf {\bibinfo {volume} {79}},\ \bibinfo
  {pages} {054002} (\bibinfo {year} {2009})},\ \Eprint
  {https://arxiv.org/abs/0811.2105} {arXiv:0811.2105 [hep-ph]} \BibitemShut
  {NoStop}%
\bibitem [{\citenamefont {Han}\ \emph {et~al.}(2010)\citenamefont {Han},
  \citenamefont {Lewis},\ and\ \citenamefont {Liu}}]{Han:2010rf}%
  \BibitemOpen
  \bibfield  {author} {\bibinfo {author} {\bibfnamefont {T.}~\bibnamefont
  {Han}}, \bibinfo {author} {\bibfnamefont {I.}~\bibnamefont {Lewis}},\ and\
  \bibinfo {author} {\bibfnamefont {Z.}~\bibnamefont {Liu}},\ }\href
  {https://doi.org/10.1007/JHEP12(2010)085} {\bibfield  {journal} {\bibinfo
  {journal} {J. High Energy Phys.}\ }\textbf {\bibinfo {volume} {12}},\
  \bibinfo {pages} {085}},\ \Eprint {https://arxiv.org/abs/1010.4309}
  {arXiv:1010.4309 [hep-ph]} \BibitemShut {NoStop}%
\bibitem [{\citenamefont {Carpenter}\ \emph
  {et~al.}(2022{\natexlab{b}})\citenamefont {Carpenter}, \citenamefont
  {Murphy},\ and\ \citenamefont {Tait}}]{Carpenter:2021rkl}%
  \BibitemOpen
  \bibfield  {author} {\bibinfo {author} {\bibfnamefont {L.~M.}\ \bibnamefont
  {Carpenter}}, \bibinfo {author} {\bibfnamefont {T.}~\bibnamefont {Murphy}},\
  and\ \bibinfo {author} {\bibfnamefont {T.~M.~P.}\ \bibnamefont {Tait}},\
  }\href {https://doi.org/10.1103/PhysRevD.105.035014} {\bibfield  {journal}
  {\bibinfo  {journal} {Phys. Rev. D}\ }\textbf {\bibinfo {volume} {105}},\
  \bibinfo {pages} {035014} (\bibinfo {year} {2022}{\natexlab{b}})},\ \Eprint
  {https://arxiv.org/abs/2110.11359} {arXiv:2110.11359 [hep-ph]} \BibitemShut
  {NoStop}%
\bibitem [{\citenamefont {Flacher}\ \emph {et~al.}(2009)\citenamefont
  {Flacher}, \citenamefont {Goebel}, \citenamefont {Haller}, \citenamefont
  {Hocker}, \citenamefont {Monig},\ and\ \citenamefont
  {Stelzer}}]{Flacher:2008zq}%
  \BibitemOpen
  \bibfield  {author} {\bibinfo {author} {\bibfnamefont {H.}~\bibnamefont
  {Flacher}}, \bibinfo {author} {\bibfnamefont {M.}~\bibnamefont {Goebel}},
  \bibinfo {author} {\bibfnamefont {J.}~\bibnamefont {Haller}}, \bibinfo
  {author} {\bibfnamefont {A.}~\bibnamefont {Hocker}}, \bibinfo {author}
  {\bibfnamefont {K.}~\bibnamefont {Monig}},\ and\ \bibinfo {author}
  {\bibfnamefont {J.}~\bibnamefont {Stelzer}},\ }\href
  {https://doi.org/10.1140/epjc/s10052-009-0966-6} {\bibfield  {journal}
  {\bibinfo  {journal} {Eur. Phys. J. C}\ }\textbf {\bibinfo {volume} {60}},\
  \bibinfo {pages} {543} (\bibinfo {year} {2009})},\ \bibinfo {note} {[Erratum:
  Eur.Phys.J.C 71, 1718 (2011)]},\ \Eprint {https://arxiv.org/abs/0811.0009}
  {arXiv:0811.0009 [hep-ph]} \BibitemShut {NoStop}%
\bibitem [{\citenamefont {Baak}\ \emph {et~al.}(2012)\citenamefont {Baak},
  \citenamefont {Goebel}, \citenamefont {Haller}, \citenamefont {Hoecker},
  \citenamefont {Ludwig}, \citenamefont {Moenig}, \citenamefont {Schott},\ and\
  \citenamefont {Stelzer}}]{Baak:2011ze}%
  \BibitemOpen
  \bibfield  {author} {\bibinfo {author} {\bibfnamefont {M.}~\bibnamefont
  {Baak}}, \bibinfo {author} {\bibfnamefont {M.}~\bibnamefont {Goebel}},
  \bibinfo {author} {\bibfnamefont {J.}~\bibnamefont {Haller}}, \bibinfo
  {author} {\bibfnamefont {A.}~\bibnamefont {Hoecker}}, \bibinfo {author}
  {\bibfnamefont {D.}~\bibnamefont {Ludwig}}, \bibinfo {author} {\bibfnamefont
  {K.}~\bibnamefont {Moenig}}, \bibinfo {author} {\bibfnamefont
  {M.}~\bibnamefont {Schott}},\ and\ \bibinfo {author} {\bibfnamefont
  {J.}~\bibnamefont {Stelzer}},\ }\href
  {https://doi.org/10.1140/epjc/s10052-012-2003-4} {\bibfield  {journal}
  {\bibinfo  {journal} {Eur. Phys. J. C}\ }\textbf {\bibinfo {volume} {72}},\
  \bibinfo {pages} {2003} (\bibinfo {year} {2012})},\ \Eprint
  {https://arxiv.org/abs/1107.0975} {arXiv:1107.0975 [hep-ph]} \BibitemShut
  {NoStop}%
\bibitem [{\citenamefont {Baak}\ \emph {et~al.}(2014)\citenamefont {Baak},
  \citenamefont {C\'uth}, \citenamefont {Haller}, \citenamefont {Hoecker},
  \citenamefont {Kogler}, \citenamefont {M\"onig}, \citenamefont {Schott},\
  and\ \citenamefont {Stelzer}}]{Baak:2014ora}%
  \BibitemOpen
  \bibfield  {author} {\bibinfo {author} {\bibfnamefont {M.}~\bibnamefont
  {Baak}}, \bibinfo {author} {\bibfnamefont {J.}~\bibnamefont {C\'uth}},
  \bibinfo {author} {\bibfnamefont {J.}~\bibnamefont {Haller}}, \bibinfo
  {author} {\bibfnamefont {A.}~\bibnamefont {Hoecker}}, \bibinfo {author}
  {\bibfnamefont {R.}~\bibnamefont {Kogler}}, \bibinfo {author} {\bibfnamefont
  {K.}~\bibnamefont {M\"onig}}, \bibinfo {author} {\bibfnamefont
  {M.}~\bibnamefont {Schott}},\ and\ \bibinfo {author} {\bibfnamefont
  {J.}~\bibnamefont {Stelzer}} (\bibinfo {collaboration} {Gfitter Group}),\
  }\href {https://doi.org/10.1140/epjc/s10052-014-3046-5} {\bibfield  {journal}
  {\bibinfo  {journal} {Eur. Phys. J. C}\ }\textbf {\bibinfo {volume} {74}},\
  \bibinfo {pages} {3046} (\bibinfo {year} {2014})},\ \Eprint
  {https://arxiv.org/abs/1407.3792} {arXiv:1407.3792 [hep-ph]} \BibitemShut
  {NoStop}%
\bibitem [{\citenamefont {Burgess}\ \emph {et~al.}(1994)\citenamefont
  {Burgess}, \citenamefont {Godfrey}, \citenamefont {Konig}, \citenamefont
  {London},\ and\ \citenamefont {Maksymyk}}]{Burgess:1993mg}%
  \BibitemOpen
  \bibfield  {author} {\bibinfo {author} {\bibfnamefont {C.~P.}\ \bibnamefont
  {Burgess}}, \bibinfo {author} {\bibfnamefont {S.}~\bibnamefont {Godfrey}},
  \bibinfo {author} {\bibfnamefont {H.}~\bibnamefont {Konig}}, \bibinfo
  {author} {\bibfnamefont {D.}~\bibnamefont {London}},\ and\ \bibinfo {author}
  {\bibfnamefont {I.}~\bibnamefont {Maksymyk}},\ }\href
  {https://doi.org/10.1016/0370-2693(94)91322-6} {\bibfield  {journal}
  {\bibinfo  {journal} {Phys. Lett. B}\ }\textbf {\bibinfo {volume} {326}},\
  \bibinfo {pages} {276} (\bibinfo {year} {1994})},\ \Eprint
  {https://arxiv.org/abs/hep-ph/9307337} {arXiv:hep-ph/9307337} \BibitemShut
  {NoStop}%
\bibitem [{\citenamefont {Grzadkowski}\ \emph {et~al.}(2010)\citenamefont
  {Grzadkowski}, \citenamefont {Iskrzyński}, \citenamefont {Misiak},\ and\
  \citenamefont {Rosiek}}]{Grzadkowski_2010}%
  \BibitemOpen
  \bibfield  {author} {\bibinfo {author} {\bibfnamefont {B.}~\bibnamefont
  {Grzadkowski}}, \bibinfo {author} {\bibfnamefont {M.}~\bibnamefont
  {Iskrzyński}}, \bibinfo {author} {\bibfnamefont {M.}~\bibnamefont
  {Misiak}},\ and\ \bibinfo {author} {\bibfnamefont {J.}~\bibnamefont
  {Rosiek}},\ }\href {https://doi.org/10.1007/jhep10(2010)085} {\bibfield
  {journal} {\bibinfo  {journal} {J. High Energy Phys.}\ }\textbf {\bibinfo
  {volume} {10}},\ \bibinfo {pages} {085}}\BibitemShut {NoStop}%
\bibitem [{\citenamefont {Brivio}\ \emph {et~al.}(2017)\citenamefont {Brivio},
  \citenamefont {Jiang},\ and\ \citenamefont {Trott}}]{Brivio_2017}%
  \BibitemOpen
  \bibfield  {author} {\bibinfo {author} {\bibfnamefont {I.}~\bibnamefont
  {Brivio}}, \bibinfo {author} {\bibfnamefont {Y.}~\bibnamefont {Jiang}},\ and\
  \bibinfo {author} {\bibfnamefont {M.}~\bibnamefont {Trott}},\ }\href
  {https://doi.org/10.1007/jhep12(2017)070} {\bibfield  {journal} {\bibinfo
  {journal} {J. High Energy Phys.}\ }\textbf {\bibinfo {volume} {12}},\
  \bibinfo {pages} {070}}\BibitemShut {NoStop}%
\bibitem [{\citenamefont {Murphy}(2020)}]{Murphy_2020}%
  \BibitemOpen
  \bibfield  {author} {\bibinfo {author} {\bibfnamefont {C.~W.}\ \bibnamefont
  {Murphy}},\ }\href {https://doi.org/10.1007/jhep10(2020)174} {\bibfield
  {journal} {\bibinfo  {journal} {J. High Energy Phys.}\ }\textbf {\bibinfo
  {volume} {10}},\ \bibinfo {pages} {174}}\BibitemShut {NoStop}%
\bibitem [{\citenamefont {Han}\ and\ \citenamefont {Skiba}(2005)}]{Han:2004az}%
  \BibitemOpen
  \bibfield  {author} {\bibinfo {author} {\bibfnamefont {Z.}~\bibnamefont
  {Han}}\ and\ \bibinfo {author} {\bibfnamefont {W.}~\bibnamefont {Skiba}},\
  }\href {https://doi.org/10.1103/PhysRevD.71.075009} {\bibfield  {journal}
  {\bibinfo  {journal} {Phys. Rev. D}\ }\textbf {\bibinfo {volume} {71}},\
  \bibinfo {pages} {075009} (\bibinfo {year} {2005})},\ \Eprint
  {https://arxiv.org/abs/hep-ph/0412166} {arXiv:hep-ph/0412166} \BibitemShut
  {NoStop}%
\bibitem [{\citenamefont {Han}(2008)}]{Han:2008es}%
  \BibitemOpen
  \bibfield  {author} {\bibinfo {author} {\bibfnamefont {Z.}~\bibnamefont
  {Han}},\ }\href {https://doi.org/10.1142/S0217751X08041384} {\bibfield
  {journal} {\bibinfo  {journal} {Int. J. Mod. Phys. A}\ }\textbf {\bibinfo
  {volume} {23}},\ \bibinfo {pages} {2653} (\bibinfo {year} {2008})},\ \Eprint
  {https://arxiv.org/abs/0807.0490} {arXiv:0807.0490 [hep-ph]} \BibitemShut
  {NoStop}%
\bibitem [{\citenamefont {Grinstein}\ and\ \citenamefont
  {Wise}(1991)}]{GRINSTEIN1991326}%
  \BibitemOpen
  \bibfield  {author} {\bibinfo {author} {\bibfnamefont {B.}~\bibnamefont
  {Grinstein}}\ and\ \bibinfo {author} {\bibfnamefont {M.~B.}\ \bibnamefont
  {Wise}},\ }\href
  {https://doi.org/https://doi.org/10.1016/0370-2693(91)90061-T} {\bibfield
  {journal} {\bibinfo  {journal} {Physics Letters B}\ }\textbf {\bibinfo
  {volume} {265}},\ \bibinfo {pages} {326} (\bibinfo {year}
  {1991})}\BibitemShut {NoStop}%
\bibitem [{\citenamefont {Martin}(1998)}]{Martin:1997sp}%
  \BibitemOpen
  \bibfield  {author} {\bibinfo {author} {\bibfnamefont {S.~P.}\ \bibnamefont
  {Martin}},\ }\href@noop {} {\bibfield  {journal} {\bibinfo  {journal} {Adv.
  Ser. Direct. High Energy Phys.}\ ,\ \bibinfo {pages} {1}} (\bibinfo {year}
  {1998})}\BibitemShut {NoStop}%
\bibitem [{\citenamefont {Lavoura}\ and\ \citenamefont
  {Li}(1994)}]{Lavoura:1993nq}%
  \BibitemOpen
  \bibfield  {author} {\bibinfo {author} {\bibfnamefont {L.}~\bibnamefont
  {Lavoura}}\ and\ \bibinfo {author} {\bibfnamefont {L.-F.}\ \bibnamefont
  {Li}},\ }\href {https://doi.org/10.1103/PhysRevD.49.1409} {\bibfield
  {journal} {\bibinfo  {journal} {Phys. Rev. D}\ }\textbf {\bibinfo {volume}
  {49}},\ \bibinfo {pages} {1409} (\bibinfo {year} {1994})},\ \Eprint
  {https://arxiv.org/abs/hep-ph/9309262} {arXiv:hep-ph/9309262} \BibitemShut
  {NoStop}%
\bibitem [{\citenamefont {Manohar}\ and\ \citenamefont
  {Wise}(2006{\natexlab{a}})}]{Manohar:2006ga}%
  \BibitemOpen
  \bibfield  {author} {\bibinfo {author} {\bibfnamefont {A.~V.}\ \bibnamefont
  {Manohar}}\ and\ \bibinfo {author} {\bibfnamefont {M.~B.}\ \bibnamefont
  {Wise}},\ }\href {https://doi.org/10.1103/PhysRevD.74.035009} {\bibfield
  {journal} {\bibinfo  {journal} {Phys. Rev. D}\ }\textbf {\bibinfo {volume}
  {74}},\ \bibinfo {pages} {035009} (\bibinfo {year} {2006}{\natexlab{a}})},\
  \Eprint {https://arxiv.org/abs/hep-ph/0606172} {arXiv:hep-ph/0606172}
  \BibitemShut {NoStop}%
\bibitem [{\citenamefont {Blank}\ and\ \citenamefont
  {Hollik}(1998)}]{Blank:1997qa}%
  \BibitemOpen
  \bibfield  {author} {\bibinfo {author} {\bibfnamefont {T.}~\bibnamefont
  {Blank}}\ and\ \bibinfo {author} {\bibfnamefont {W.}~\bibnamefont {Hollik}},\
  }\href {https://doi.org/10.1016/S0550-3213(97)00785-2} {\bibfield  {journal}
  {\bibinfo  {journal} {Nucl. Phys. B}\ }\textbf {\bibinfo {volume} {514}},\
  \bibinfo {pages} {113} (\bibinfo {year} {1998})},\ \Eprint
  {https://arxiv.org/abs/hep-ph/9703392} {arXiv:hep-ph/9703392} \BibitemShut
  {NoStop}%
\bibitem [{\citenamefont {Khandker}\ \emph {et~al.}(2012)\citenamefont
  {Khandker}, \citenamefont {Li},\ and\ \citenamefont
  {Skiba}}]{Khandker:2012zu}%
  \BibitemOpen
  \bibfield  {author} {\bibinfo {author} {\bibfnamefont {Z.~U.}\ \bibnamefont
  {Khandker}}, \bibinfo {author} {\bibfnamefont {D.}~\bibnamefont {Li}},\ and\
  \bibinfo {author} {\bibfnamefont {W.}~\bibnamefont {Skiba}},\ }\href
  {https://doi.org/10.1103/PhysRevD.86.015006} {\bibfield  {journal} {\bibinfo
  {journal} {Phys. Rev. D}\ }\textbf {\bibinfo {volume} {86}},\ \bibinfo
  {pages} {015006} (\bibinfo {year} {2012})},\ \Eprint
  {https://arxiv.org/abs/1201.4383} {arXiv:1201.4383 [hep-ph]} \BibitemShut
  {NoStop}%
\bibitem [{\citenamefont {Peskin}\ and\ \citenamefont
  {Schroeder}(1995)}]{PSQFT}%
  \BibitemOpen
  \bibfield  {author} {\bibinfo {author} {\bibfnamefont {M.~E.}\ \bibnamefont
  {Peskin}}\ and\ \bibinfo {author} {\bibfnamefont {D.~V.}\ \bibnamefont
  {Schroeder}},\ }\href@noop {} {\emph {\bibinfo {title} {{An Introduction to
  Quantum Field Theory}}}}\ (\bibinfo  {publisher} {Addison Wesley
  Publishing},\ \bibinfo {year} {1995})\BibitemShut {NoStop}%
\bibitem [{\citenamefont {Bethke}(2007)}]{BETHKE2007351}%
  \BibitemOpen
  \bibfield  {author} {\bibinfo {author} {\bibfnamefont {S.}~\bibnamefont
  {Bethke}},\ }\href
  {https://doi.org/https://doi.org/10.1016/j.ppnp.2006.06.001} {\bibfield
  {journal} {\bibinfo  {journal} {Progress in Particle and Nuclear Physics}\
  }\textbf {\bibinfo {volume} {58}},\ \bibinfo {pages} {351} (\bibinfo {year}
  {2007})}\BibitemShut {NoStop}%
\bibitem [{\citenamefont {Carpenter}\ \emph {et~al.}(2021)\citenamefont
  {Carpenter}, \citenamefont {Smylie}, \citenamefont {Ramirez}, \citenamefont
  {McDowell},\ and\ \citenamefont {Whiteson}}]{Carpenter:2021gpl}%
  \BibitemOpen
  \bibfield  {author} {\bibinfo {author} {\bibfnamefont {L.~M.}\ \bibnamefont
  {Carpenter}}, \bibinfo {author} {\bibfnamefont {M.~J.}\ \bibnamefont
  {Smylie}}, \bibinfo {author} {\bibfnamefont {J.~M.~C.}\ \bibnamefont
  {Ramirez}}, \bibinfo {author} {\bibfnamefont {C.}~\bibnamefont {McDowell}},\
  and\ \bibinfo {author} {\bibfnamefont {D.}~\bibnamefont {Whiteson}},\
  }\Eprint {https://arxiv.org/abs/2112.00137} {arXiv:2112.00137 [hep-ph]}
  (\bibinfo {year} {2021})\BibitemShut {NoStop}%
\bibitem [{\citenamefont {Christensen}\ and\ \citenamefont
  {Duhr}(2009)}]{FR_OG}%
  \BibitemOpen
  \bibfield  {author} {\bibinfo {author} {\bibfnamefont {N.~D.}\ \bibnamefont
  {Christensen}}\ and\ \bibinfo {author} {\bibfnamefont {C.}~\bibnamefont
  {Duhr}},\ }\href {https://doi.org/10.1016/j.cpc.2009.02.018} {\bibfield
  {journal} {\bibinfo  {journal} {Comput. Phys. Commun.}\ }\textbf {\bibinfo
  {volume} {180}},\ \bibinfo {pages} {1614–1641} (\bibinfo {year}
  {2009})}\BibitemShut {NoStop}%
\bibitem [{\citenamefont {Alloul}\ \emph {et~al.}(2014)\citenamefont {Alloul},
  \citenamefont {Christensen}, \citenamefont {Degrande}, \citenamefont {Duhr},\
  and\ \citenamefont {Fuks}}]{FR_2}%
  \BibitemOpen
  \bibfield  {author} {\bibinfo {author} {\bibfnamefont {A.}~\bibnamefont
  {Alloul}}, \bibinfo {author} {\bibfnamefont {N.~D.}\ \bibnamefont
  {Christensen}}, \bibinfo {author} {\bibfnamefont {C.}~\bibnamefont
  {Degrande}}, \bibinfo {author} {\bibfnamefont {C.}~\bibnamefont {Duhr}},\
  and\ \bibinfo {author} {\bibfnamefont {B.}~\bibnamefont {Fuks}},\ }\href
  {https://doi.org/10.1016/j.cpc.2014.04.012} {\bibfield  {journal} {\bibinfo
  {journal} {Comput. Phys. Commun.}\ }\textbf {\bibinfo {volume} {185}},\
  \bibinfo {pages} {2250–2300} (\bibinfo {year} {2014})}\BibitemShut
  {NoStop}%
\bibitem [{Mat(2020)}]{Mathematica}%
  \BibitemOpen
  \href {https://www.wolfram.com/mathematica} {\bibinfo {title} {Wolfram
  {R}esearch, {I}nc., \textsc{Mathematica}$^{\copyright}$, {V}ersion 12.0}}
  (\bibinfo {year} {2020})\BibitemShut {NoStop}%
\bibitem [{\citenamefont {Hahn}(2001)}]{FA}%
  \BibitemOpen
  \bibfield  {author} {\bibinfo {author} {\bibfnamefont {T.}~\bibnamefont
  {Hahn}},\ }\href {https://doi.org/10.1016/s0010-4655(01)00290-9} {\bibfield
  {journal} {\bibinfo  {journal} {Comput. Phys. Commun.}\ }\textbf {\bibinfo
  {volume} {140}},\ \bibinfo {pages} {418–431} (\bibinfo {year}
  {2001})}\BibitemShut {NoStop}%
\bibitem [{\citenamefont {Shtabovenko}(2017)}]{feynhelpers}%
  \BibitemOpen
  \bibfield  {author} {\bibinfo {author} {\bibfnamefont {V.}~\bibnamefont
  {Shtabovenko}},\ }\href {https://doi.org/10.1016/j.cpc.2017.04.014}
  {\bibfield  {journal} {\bibinfo  {journal} {Comput. Phys. Commun.}\ }\textbf
  {\bibinfo {volume} {218}},\ \bibinfo {pages} {48–65} (\bibinfo {year}
  {2017})}\BibitemShut {NoStop}%
\bibitem [{\citenamefont {Mertig}\ \emph {et~al.}(1991)\citenamefont {Mertig},
  \citenamefont {Böhm},\ and\ \citenamefont {Denner}}]{MERTIG1991345}%
  \BibitemOpen
  \bibfield  {author} {\bibinfo {author} {\bibfnamefont {R.}~\bibnamefont
  {Mertig}}, \bibinfo {author} {\bibfnamefont {M.}~\bibnamefont {Böhm}},\ and\
  \bibinfo {author} {\bibfnamefont {A.}~\bibnamefont {Denner}},\ }\href
  {https://doi.org/https://doi.org/10.1016/0010-4655(91)90130-D} {\bibfield
  {journal} {\bibinfo  {journal} {Comput. Phys. Commun.}\ }\textbf {\bibinfo
  {volume} {64}},\ \bibinfo {pages} {345} (\bibinfo {year} {1991})}\BibitemShut
  {NoStop}%
\bibitem [{\citenamefont {Shtabovenko}\ \emph {et~al.}(2016)\citenamefont
  {Shtabovenko}, \citenamefont {Mertig},\ and\ \citenamefont
  {Orellana}}]{FC_9.0}%
  \BibitemOpen
  \bibfield  {author} {\bibinfo {author} {\bibfnamefont {V.}~\bibnamefont
  {Shtabovenko}}, \bibinfo {author} {\bibfnamefont {R.}~\bibnamefont
  {Mertig}},\ and\ \bibinfo {author} {\bibfnamefont {F.}~\bibnamefont
  {Orellana}},\ }\href {https://doi.org/10.1016/j.cpc.2016.06.008} {\bibfield
  {journal} {\bibinfo  {journal} {Comput. Phys. Commun.}\ }\textbf {\bibinfo
  {volume} {207}},\ \bibinfo {pages} {432–444} (\bibinfo {year}
  {2016})}\BibitemShut {NoStop}%
\bibitem [{\citenamefont {Shtabovenko}\ \emph {et~al.}(2020)\citenamefont
  {Shtabovenko}, \citenamefont {Mertig},\ and\ \citenamefont
  {Orellana}}]{FC_9.3.0}%
  \BibitemOpen
  \bibfield  {author} {\bibinfo {author} {\bibfnamefont {V.}~\bibnamefont
  {Shtabovenko}}, \bibinfo {author} {\bibfnamefont {R.}~\bibnamefont
  {Mertig}},\ and\ \bibinfo {author} {\bibfnamefont {F.}~\bibnamefont
  {Orellana}},\ }\href {https://doi.org/10.1016/j.cpc.2020.107478} {\bibfield
  {journal} {\bibinfo  {journal} {Comput. Phys. Commun.}\ }\textbf {\bibinfo
  {volume} {256}},\ \bibinfo {pages} {107478} (\bibinfo {year}
  {2020})}\BibitemShut {NoStop}%
\bibitem [{\citenamefont {Patel}(2017)}]{Patel:2017px}%
  \BibitemOpen
  \bibfield  {author} {\bibinfo {author} {\bibfnamefont {H.}~\bibnamefont
  {Patel}},\ }\href {https://doi.org/10.1016/j.cpc.2017.04.015} {\bibfield
  {journal} {\bibinfo  {journal} {Comput. Phys. Commun.}\ }\textbf {\bibinfo
  {volume} {218}},\ \bibinfo {pages} {66} (\bibinfo {year} {2017})}\BibitemShut
  {NoStop}%
\bibitem [{\citenamefont {Aaboud}\ \emph
  {et~al.}(2018{\natexlab{b}})\citenamefont {Aaboud} \emph
  {et~al.}}]{ATLAS:2017jnp}%
  \BibitemOpen
  \bibfield  {author} {\bibinfo {author} {\bibfnamefont {M.}~\bibnamefont
  {Aaboud}} \emph {et~al.} (\bibinfo {collaboration} {ATLAS}),\ }\href
  {https://doi.org/10.1140/epjc/s10052-018-5693-4} {\bibfield  {journal}
  {\bibinfo  {journal} {Eur. Phys. J. C}\ }\textbf {\bibinfo {volume} {78}},\
  \bibinfo {pages} {250} (\bibinfo {year} {2018}{\natexlab{b}})},\ \Eprint
  {https://arxiv.org/abs/1710.07171} {arXiv:1710.07171 [hep-ex]} \BibitemShut
  {NoStop}%
\bibitem [{\citenamefont {Sirunyan}\ \emph {et~al.}(2018)\citenamefont
  {Sirunyan} \emph {et~al.}}]{CMS:2018mgb}%
  \BibitemOpen
  \bibfield  {author} {\bibinfo {author} {\bibfnamefont {A.~M.}\ \bibnamefont
  {Sirunyan}} \emph {et~al.} (\bibinfo {collaboration} {CMS}),\ }\href
  {https://doi.org/10.1007/JHEP08(2018)130} {\bibfield  {journal} {\bibinfo
  {journal} {J. High Energy Phys.}\ }\textbf {\bibinfo {volume} {08}},\
  \bibinfo {pages} {130}},\ \Eprint {https://arxiv.org/abs/1806.00843}
  {arXiv:1806.00843 [hep-ex]} \BibitemShut {NoStop}%
\bibitem [{\citenamefont {Manohar}\ and\ \citenamefont
  {Wise}(2006{\natexlab{b}})}]{Manohar:2006gz}%
  \BibitemOpen
  \bibfield  {author} {\bibinfo {author} {\bibfnamefont {A.~V.}\ \bibnamefont
  {Manohar}}\ and\ \bibinfo {author} {\bibfnamefont {M.~B.}\ \bibnamefont
  {Wise}},\ }\href {https://doi.org/10.1016/j.physletb.2006.03.030} {\bibfield
  {journal} {\bibinfo  {journal} {Phys. Lett. B}\ }\textbf {\bibinfo {volume}
  {636}},\ \bibinfo {pages} {107} (\bibinfo {year} {2006}{\natexlab{b}})},\
  \Eprint {https://arxiv.org/abs/hep-ph/0601212} {arXiv:hep-ph/0601212}
  \BibitemShut {NoStop}%
\bibitem [{\citenamefont {Hayreter}\ and\ \citenamefont
  {Valencia}(2017)}]{Hayreter:2017wra}%
  \BibitemOpen
  \bibfield  {author} {\bibinfo {author} {\bibfnamefont {A.}~\bibnamefont
  {Hayreter}}\ and\ \bibinfo {author} {\bibfnamefont {G.}~\bibnamefont
  {Valencia}},\ }\href {https://doi.org/10.1103/PhysRevD.96.035004} {\bibfield
  {journal} {\bibinfo  {journal} {Phys. Rev. D}\ }\textbf {\bibinfo {volume}
  {96}},\ \bibinfo {pages} {035004} (\bibinfo {year} {2017})},\ \Eprint
  {https://arxiv.org/abs/1703.04164} {arXiv:1703.04164 [hep-ph]} \BibitemShut
  {NoStop}%
\bibitem [{\citenamefont {Cacciapaglia}\ \emph {et~al.}(2020)\citenamefont
  {Cacciapaglia}, \citenamefont {Deandrea}, \citenamefont {Flacke},\ and\
  \citenamefont {Iyer}}]{Cacciapaglia:2020vyf}%
  \BibitemOpen
  \bibfield  {author} {\bibinfo {author} {\bibfnamefont {G.}~\bibnamefont
  {Cacciapaglia}}, \bibinfo {author} {\bibfnamefont {A.}~\bibnamefont
  {Deandrea}}, \bibinfo {author} {\bibfnamefont {T.}~\bibnamefont {Flacke}},\
  and\ \bibinfo {author} {\bibfnamefont {A.~M.}\ \bibnamefont {Iyer}},\ }\href
  {https://doi.org/10.1007/JHEP05(2020)027} {\bibfield  {journal} {\bibinfo
  {journal} {J. High Energy Phys.}\ }\textbf {\bibinfo {volume} {05}},\
  \bibinfo {pages} {027}},\ \Eprint {https://arxiv.org/abs/2002.01474}
  {arXiv:2002.01474 [hep-ph]} \BibitemShut {NoStop}%
\bibitem [{\citenamefont {Miralles}\ and\ \citenamefont
  {Pich}(2019)}]{Miralles:2019uzg}%
  \BibitemOpen
  \bibfield  {author} {\bibinfo {author} {\bibfnamefont {V.}~\bibnamefont
  {Miralles}}\ and\ \bibinfo {author} {\bibfnamefont {A.}~\bibnamefont
  {Pich}},\ }\href {https://doi.org/10.1103/PhysRevD.100.115042} {\bibfield
  {journal} {\bibinfo  {journal} {Phys. Rev. D}\ }\textbf {\bibinfo {volume}
  {100}},\ \bibinfo {pages} {115042} (\bibinfo {year} {2019})},\ \Eprint
  {https://arxiv.org/abs/1910.07947} {arXiv:1910.07947 [hep-ph]} \BibitemShut
  {NoStop}%
\bibitem [{\citenamefont {Eberhardt}\ \emph {et~al.}(2021)\citenamefont
  {Eberhardt}, \citenamefont {Miralles},\ and\ \citenamefont
  {Pich}}]{Eberhardt:2021ebh}%
  \BibitemOpen
  \bibfield  {author} {\bibinfo {author} {\bibfnamefont {O.}~\bibnamefont
  {Eberhardt}}, \bibinfo {author} {\bibfnamefont {V.}~\bibnamefont
  {Miralles}},\ and\ \bibinfo {author} {\bibfnamefont {A.}~\bibnamefont
  {Pich}},\ }\href {https://doi.org/10.1007/JHEP10(2021)123} {\bibfield
  {journal} {\bibinfo  {journal} {J. High Energy Phys.}\ }\textbf {\bibinfo
  {volume} {10}},\ \bibinfo {pages} {123}},\ \Eprint
  {https://arxiv.org/abs/2106.12235} {arXiv:2106.12235 [hep-ph]} \BibitemShut
  {NoStop}%
\bibitem [{\citenamefont {Boughezal}\ and\ \citenamefont
  {Petriello}(2010)}]{Boughezal:2010ry}%
  \BibitemOpen
  \bibfield  {author} {\bibinfo {author} {\bibfnamefont {R.}~\bibnamefont
  {Boughezal}}\ and\ \bibinfo {author} {\bibfnamefont {F.}~\bibnamefont
  {Petriello}},\ }\href {https://doi.org/10.1103/PhysRevD.81.114033} {\bibfield
   {journal} {\bibinfo  {journal} {Phys. Rev. D}\ }\textbf {\bibinfo {volume}
  {81}},\ \bibinfo {pages} {114033} (\bibinfo {year} {2010})},\ \Eprint
  {https://arxiv.org/abs/1003.2046} {arXiv:1003.2046 [hep-ph]} \BibitemShut
  {NoStop}%
\bibitem [{\citenamefont {Schael}\ \emph {et~al.}(2006)\citenamefont {Schael}
  \emph {et~al.}}]{ALEPH:2005ab}%
  \BibitemOpen
  \bibfield  {author} {\bibinfo {author} {\bibfnamefont {S.}~\bibnamefont
  {Schael}} \emph {et~al.} (\bibinfo {collaboration} {ALEPH, DELPHI, L3, OPAL,
  SLD, LEP Electroweak Working Group, SLD Electroweak Group, SLD Heavy Flavour
  Group}),\ }\href {https://doi.org/10.1016/j.physrep.2005.12.006} {\bibfield
  {journal} {\bibinfo  {journal} {Phys. Rept.}\ }\textbf {\bibinfo {volume}
  {427}},\ \bibinfo {pages} {257} (\bibinfo {year} {2006})},\ \Eprint
  {https://arxiv.org/abs/hep-ex/0509008} {arXiv:hep-ex/0509008} \BibitemShut
  {NoStop}%
\bibitem [{\citenamefont {Aguilar-Arevalo}\ \emph {et~al.}(2018)\citenamefont
  {Aguilar-Arevalo} \emph {et~al.}}]{miniboone_2018}%
  \BibitemOpen
  \bibfield  {author} {\bibinfo {author} {\bibfnamefont {A.}~\bibnamefont
  {Aguilar-Arevalo}} \emph {et~al.} (\bibinfo {collaboration} {MiniBooNE
  Collaboration}),\ }\href {https://doi.org/10.1103/physrevlett.121.221801}
  {\bibfield  {journal} {\bibinfo  {journal} {Phys. Rev. Lett.}\ }\textbf
  {\bibinfo {volume} {121}},\ \bibinfo {pages} {221801} (\bibinfo {year}
  {2018})}\BibitemShut {NoStop}%
\end{thebibliography}%
